\documentclass[prd,nofootinbib,superscriptaddress,eqsecnum,tightenlines,11pt]{revtex4}

\usepackage{amsfonts}
\usepackage{amsmath}
\usepackage{amssymb}
\usepackage{hyperref}
\usepackage{stmaryrd}
\usepackage{color}

\def\be{\begin{equation}}
\def\ee{\end{equation}}
\def\bes{\begin{equation*}}
\def\ees{\end{equation*}}
\def\ba{\begin{eqnarray}}
\def\ea{\end{eqnarray}}

\def\eps{\varepsilon}
\def\lp{l_\text{Pl}}

\def\tr{\text{tr}}

\def\de{\mathrm{d}}

\newcommand{\f}{\frac}
\newcommand{\lb}{\big\lbrace}
\newcommand{\rb}{\big\rbrace}
\newcommand{\SU}{\text{SU}}
\newcommand{\DSU}{\text{DSU}}
\newcommand{\ISU}{\text{ISU}}
\newcommand{\SO}{\text{SO}}

\newcommand{\SL}{\text{SL}}
\newcommand{\isu}{\mathfrak{isu}}
\newcommand{\su}{\mathfrak{su}}
\renewcommand{\sl}{\mathfrak{sl}}
\newcommand{\so}{\mathfrak{so}}

\renewcommand\time{\,{\scriptstyle{\times}}\,}

\begin{document}

\title{Testing the role of the Barbero-Immirzi parameter and\\ the choice of connection in Loop Quantum Gravity}

\author{Jibril Ben Achour}
\affiliation{Laboratoire APC -- Astroparticule et Cosmologie, Universit\'e Paris Diderot Paris 7, 75013 Paris, France}
\author{Marc Geiller}
\affiliation{Institute for Gravitation and the Cosmos \& Physics Department, Penn State, University Park, PA 16802, U.S.A.}
\author{Karim Noui}
\affiliation{Laboratoire de Math\'ematiques et Physique Th\'eorique, Universit\'e Fran\c cois Rabelais, Parc de Grandmont, 37200 Tours, France}
\affiliation{Laboratoire APC -- Astroparticule et Cosmologie, Universit\'e Paris Diderot Paris 7, 75013 Paris, France}
\author{Chao Yu}
\affiliation{\'ENS de Lyon, 46, all\'ee d'Italie, 69007 Lyon, France\bigskip}

\begin{abstract}
We study the role of the Barbero-Immirzi parameter $\gamma$ and the choice of connection in the construction of (a symmetry-reduced version of) loop quantum gravity. We start with the four-dimensional Lorentzian Holst action that we reduce to three dimensions in a way that preserves the presence of $\gamma$. In the time gauge, the phase space of the resulting three-dimensional theory mimics exactly that of the four-dimensional one. Its quantization can be performed, and on the kinematical Hilbert space spanned by $\SU(2)$ spin network states the spectra of geometric operators are discrete and $\gamma$-dependent. However, because of the three-dimensional nature of the theory, its $\SU(2)$ Ashtekar-Barbero Hamiltonian constraint can be traded for the flatness constraint of an $\sl(2,\mathbb{C})$ connection, and we show that this latter has to satisfy a linear simplicity-like condition analogous to the one used in the construction of spin foam models. The physically relevant solution to this constraint singles out the non-compact subgroup $\SU(1,1)$, which in turn leads to the disappearance of the Barbero-Immirzi parameter and to a continuous length spectrum, in agreement with what is expected from Lorentzian three-dimensional gravity.
\end{abstract}

\maketitle

\newpage
\tableofcontents

\newpage

\section{Introduction}

\noindent Since the introduction of the $\SU(2)$ Ashtekar-Barbero variables \cite{barbero,immirzi} as a way to circumvent the problem of imposing the reality conditions in the original complex Ashtekar formulation of gravity \cite{complexashtekar}, the status of the Barbero-Immirzi parameter $\gamma$ has been debated intensively \cite{rovelli-thiemann,menamarugan,fermions1,fermions2,fermions3,taveras-yunes,fermions4,fermions5,mercuri-taveras,mercuri-gamma,mercuri-gamma2,mercuri-randono,dittrich-ryan2,GN}. The reason for this is that although $\gamma$ plays no role at the classical level, it is manifestly present in the quantum theory, for example in the spectrum of the various geometrical operators \cite{rovelli-smolin-area-volume,ashtekar-lewandowski-area,ashtekar-lewandowski-volume}. It is often stated that the family of canonical transformations labelled by the Barbero-Immirzi parameter cannot be implemented unitarily upon quantization.

From the Lagrangian point of view, $\gamma$ drops out of the classical theory (the Holst action \cite{holst}) by virtue of the Bianchi identities once the torsion-free condition is imposed (i.e. on half-shell). In the Hamiltonian formulation in terms of the Ashtekar-Barbero connection, $\gamma$ is featured in both the Poisson bracket between the phase space variables and the scalar constraint, and the mechanism responsible for its disappearance is more complicated since the covariant torsion two-form is broken into its various components, some of which correspond to the Hamiltonian evolution of the triad field. This is of course to be expected since at the Lagrangian level it is also (half of) the equations of motion that are responsible for the disappearance of the Barbero-Immirzi parameter. This observation however raises the question of the fate of $\gamma$ in the full Hamiltonian quantum theory, i.e. when the dynamics is taken into account. Although there is a well-defined and anomaly-free regularization of the quantum scalar constraint of loop quantum gravity, the physical state space is not fully characterized, and nothing is known about the fate of $\gamma$ at the physical level. From the perspective of the recent four-dimensional spin foam models (see \cite{aleSF} for a review) the Barbero-Immirzi parameter plays a priori a non-trivial (and even central) role in the definition of the dynamics of quantum gravity. However, $\gamma$ was essentially introduced in these models in order for the boundary states to match that of canonical loop quantum gravity, and there are no clear indications as to why a spin foam model would necessarily require the presence of the Barbero-Immirzi parameter in order to be well-defined.

Recently, it has been observed in the context of work on black hole entropy \cite{FGNP,BST,pranzetti} and on the asymptotic behavior of Lorentzian spin foam amplitudes \cite{BN} that the self-dual value $\gamma=\pm\mathrm{i}$ can be chosen consistently (at least in some specific calculations). This points towards the potentially underestimated role played by the original complex Ashtekar variables in the quantization of gravity, and the fact that the Barbero-Immirzi ambiguity can potentially be resolved by simply going back to the self-dual value. Of course, when working with $\gamma=\pm\mathrm{i}$ from the onset one runs into the problems of defining spin network states for a non-compact gauge group and of imposing the reality conditions. It is however relevant to try to investigate the relationship between the present formulation of $\SU(2)$ loop quantum gravity with the Barbero-Immirzi parameter, and the yet-to-be defined self-dual quantum theory.

This paper is a first step towards the investigation of this relationship. Because of the difficulties present in treating the dynamics of four-dimensional loop quantum gravity, we consider here a symmetry-reduced model which captures the essential features of the full theory. This symmetry reduction consists in imposing invariance of the four-dimensional Holst theory along a given spatial direction, which reduces the action to that of three-dimensional gravity with a Barbero-Immirzi parameter. Then, the construction of the physical Hilbert space becomes in principle possible (although we do not present it) because three-dimensional quantum gravity is an exactly soluble system \cite{Witten 3d}, and one can properly phrase the question of the fate of the Barbero-Immirzi parameter. In the time gauge, the $\SU(2)$ kinematical structure is $\gamma$-dependent just like in the four-dimensional theory. However, if one tries to rewrite the Hamiltonian constraint in the form of a flatness constraint in order to simplify its imposition at the quantum level, the connection turns out to become complex, and the associated reality conditions take the form of the linear simplicity constraint of spin foam models. Once this simplicity constraint is imposed at the classical level, we observe that the Barbero-Immirzi parameter disappears already in the kinematical quantum theory, and that the physical states are just given by that of three-dimensional $\SU(1,1)$ BF theory.

Our starting point for this study is the action for three-dimensional Lorentzian gravity with a Barbero-Immirzi parameter introduced and partially studied in \cite{GN,GN2}. This action can be obtained from a spacetime symmetry reduction of the four-dimensional Holst action, and as such its internal symmetry group is the Lorentz group $G=\SL(2,\mathbb{C})$. In \cite{GN2}, it was shown that there are two ``natural'' gauge fixings of the internal gauge group $G$ that lead to two different (but nonetheless physically equivalent) parametrizations of the phase space. When the gauge fixing is chosen in such a way that $G$ is broken into its non-compact subgroup $\SU(1,1)$, the theory reduces to $\SU(1,1)$ BF theory, and the Barbero-Immirzi parameter completely disappears from the phase space. The quantization can then be performed in principle using the combinatorial quantization scheme \cite{Alekseev:1994pa,Alekseev:1994au,BNR,Catherine}, and if one were to carry out this very technical but nonetheless well-defined procedure, the resulting physical state space would be $\gamma$-independent (for the simple reason that in this case $\gamma$ is already absent from the kinematical structure once the second class constraints are taken into account).

More interestingly, when the gauge fixing is chosen to be the three-dimensional version of the time gauge, $G$ is broken into its compact $\SU(2)$ subgroup, and the resulting Hamiltonian formulation is exactly analogous to that of the four-dimensional Ashtekar-Barbero theory. In particular, the fundamental variables in this case are the three-dimensional version $A^i_a$ of the Ashtekar-Barbero connection and its conjugate triad $E^a_i$, and their Poisson bracket as well as the scalar constraint are $\gamma$-dependent. Following the procedure of canonical loop quantization, we arrive at a well-defined kinematical Hilbert space with a discrete and $\gamma$-dependent length spectrum. The Hamiltonian constraint can then be solved in two ways. The first one is to follow the regularization procedure introduced for the definition of the four-dimensional scalar constraint, and the second one is to reformulate the theory in a BF manner at the classical level before performing its quantization. Following this second direction, we show that it is indeed possible to rewrite the three-dimensional $\SU(2)$ Ashtekar-Barbero theory as a BF theory, but at the expense of making the connection complex. Unfortunately, by doing so we then face the same difficulties as if we had started from the onset with the self-dual connection obtained by choosing $\gamma=\pm\mathrm{i}$. However, guided by the fact that in the aforementioned non-compact gauge the original action reduces to that of $\SU(1,1)$ BF theory, we argue that the reality conditions should select the subgroup $\SU(1,1)$ as the real section of the Lorentz group $\SL(2,\mathbb{C})$. This requirement is met if and only if the elements $J_i$ and $P_i$ representing the generators of infinitesimal rotations and boosts satisfy a condition analogous to the linear simplicity constraint arising in the construction of four-dimensional spin foam models. Its appearance here is not much of a surprise, since it is known also in four dimensions that when working with the Holst action (i.e. with $\gamma\in\mathbb{R}$) and using the (anti) self-dual decomposition of the variables, the reality condition is traded for the simplicity constraint \cite{wieland}. The interesting observation in our context is that this constraint admits two types of solutions.

The first one consists in modifying the action of the infinitesimal boosts $P_i$, while keeping the action of the infinitesimal rotations $J_i$ unchanged. By doing so, the action of boosts is somehow compactified, and one can show that the $\SL(2,\mathbb{C})$ connection reduces to the initial $\SU(2)$ Ashtekar-Barbero connection. The second solution (which turns out to involve the generators of $\su(1,1)$ and their complement in $\sl(2,\mathbb{C})$) does however reduces the complex connection to an $\SU(1,1)$ connection. In the EPRL and FK${}_\gamma$ spin foam models, the first solution is selected. Here, we select the second one because the resulting connection is $\su(1,1)$-valued, which is what we expect from the requirement of consistency with the non-compact gauge. Furthermore, in this case, the Barbero-Immirzi parameter
drops out of the theory and does not play any role at the physical level. Therefore, in this three-dimensional model, it seems that there is a consistent way of sending the $\SU(2)$ Ashtekar-Barbero connection to an $\SU(1,1)$ connection and to get a theory that at the end of the day does not depend on $\gamma$ anymore.

This paper is organized as follows. In section \ref{sec:classical} we introduce the Lorentzian three-dimensional Holst action and review some of its properties at the Lagrangian level. Section \ref{sec:gauge choices} takes the classical analysis a step further by studying in detail the theory in two different gauges. The first one reduces the Lorentz group to $\SU(1,1)$, and the action becomes that of $\SU(1,1)$ BF theory, while the second gauge choice leads to the $\SU(2)$ Ashtekar-Barbero phase space. Section \ref{sec:quantum theory} is devoted to the study of the quantum theory. After briefly discussing the quantization strategies available for dealing with $\SU(1,1)$ BF theory, we tackle the issue of quantizing the $\SU(2)$ theory in the time gauge. We show that the Ashtekar-Barbero constraints can be written in a BF fashion at the expense of working with a complex connection, and that there is a consistent way of imposing the requirement that the physical states be $\SU(1,1)$ spin networks, which in turn leads to the elimination of the Barbero-Immirzi parameter. We conclude with a discussion on the possibility of extending these results to the full four-dimensional theory. 

\section{Classical theory}
\label{sec:classical}

\noindent In four spacetime dimensions, the first order action for general relativity that serves as a starting point for canonical loop quantum gravity is given by \cite{holst}
\be\label{holst action}
S_\text{4D}[e,\omega]=\int_{\mathcal{M}_4}\left(\frac{1}{2}\eps_{IJKL}e^I\wedge e^J \wedge 
F^{KL}+\frac{1}{\gamma}\delta_{IJKL}e^I\wedge e^J\wedge F^{KL}\right).
\ee
The dynamical variables are the tetrad one-form fields $e^I_\mu$ and the $\sl(2,\mathbb{C})$-valued connection $\omega^{IJ}_\mu$, whose curvature is denoted by $F=\de\omega+(\omega\wedge\omega)/2$. The totally antisymmetric tensor $\eps_{IJKL}$ is the Killing form on $\sl(2,\mathbb{C})$, and $\delta_{IJKL}=(\eta_{IK}\eta_{JL}-\eta_{IL}\eta_{JK})/2$, with $\eta_{IJ}=\text{diag}(-1,1,1,1)$ the flat metric, is the other independent invariant bilinear form on $\sl(2,\mathbb{C})$ (with a suitable normalization). 

In loop quantum gravity, one chooses to work in the time gauge, which consists in breaking the $\SL(2,\mathbb{C})$ gauge group into an $\SU(2)$ maximal compact subgroup by imposing the conditions $e_a^0=0$. In this case, the canonical analysis simplifies dramatically \cite{holst,GN}, and the phase space is parametrized by an $\su(2)$-valued connection known as the Ashtekar-Barbero connection, together with its conjugate densitized triad field. The quantization then leads to a mathematically well-defined kinematical Hilbert space because of the compactness of the gauge group. At the kinematical level, the area and volume operators exhibit discrete spectra, and the Barbero-Immirzi parameter can be interpreted as a measure of the area gap in Planck units.

Without fixing the time gauge, the canonical analysis of the Holst action is quite involved, and was performed originally in \cite{alexandrov1,alexandrov2}. Once the second class constraints are taken into account, there are essentially two choices of connection that can be made. The first one corresponds to the Lorentz-covariant extension of the Ashtekar-Barbero connection. This connection is commutative with respect to the Dirac bracket, and leads to the same quantum theory as the $\SU(2)$ formulation in the time gauge \cite{alexandrov4,GLNS,GLN}. However, just like its $\SU(2)$ counterpart, the Lorentz-covariant Ashtekar-Barebro connection is not the pullback of a spacetime connection \cite{alexandrov5}. Instead, one can choose to work with the shifted connection, which can be interpreted as a spacetime connection but has the disadvantage of being non-commutative. In this case, the Hamiltonian formulation becomes completely independent of $\gamma$, which drops out of the theory as expected from the Lagrangian analysis. This strongly suggests that in this particular Lorentz-covariant formulation with the shifted connection the Barbero-Immirzi parameter will play no role at the quantum level. Unfortunately, no representation of the associated quantum algebra has ever been found (see however \cite{alexandrov6} for an attempt). It has however been argued by Alexandrov that this quantization could lead to a continuous area spectrum with no dependency on $\gamma$ \cite{alexandrov5}. This non-dependency on $\gamma$ is a completely natural thing to expect if one starts with a canonical formulation which is already $\gamma$-independent at the classical level.

Evidently, there seems to be a discrepancy between the predictions of the quantum theories based on the Ashtekar-Barbero connection (either in the $\SU(2)$ or Lorentz-covariant formulation) and the shifted connection. However, the problem is that up to now none of these derivations are fully understood. Indeed, as we have just mentioned above, the kinematical states are not even defined in the Lorentz-covariant quantization with the shifted connection due to the non-commutativity of the connection, and in the quantization in the time gauge we do not have full control over the physical Hilbert space and the geometrical operators are only defined at the kinematical level\footnote{Although it has been argued in \cite{dittrich-thiemann} that the nature of the spectra could change at the physical level, the results of \cite{giesel-thiemann,DGKL} seem to indicate that they remain discrete. As pointed out in \cite{BST,BST2}, this interpretation is tightly linked to the choice a classical time function, which in turn dictates the nature of the gauge-invariant geometric operators.}. It is nonetheless honest to say that the quantization of the $\SU(2)$ theory in the time gauge is much more advanced and mathematically well-defined, although it is very interesting and intriguing that the Lorentz-covariant theory points towards important issues concerning the status of the Barbero-Immirzi parameter and the relevance of the $\SU(2)$ Ashtekar-Barbero connection.

We are going to present a formulation of three-dimensional gravity that can help understand the tensions that we have just described. This model was originally introduced in \cite{GN2} in the context of spin foam models, and further studied in \cite{GN} in order to illustrate the interplay between the gauge-fixing of the Holst action and the role of the Barbero-Immirzi parameter. It can be obtained by a reduction of the four-dimensional Holst action to three dimensions. In this section, we will present this model in details and recall its classical properties.

\subsection{Symmetry reduction from 4 to 3 dimensions}

\noindent Starting with the four-dimensional Holst action (\ref{holst action}), we perform a spacetime compactification without reducing the internal gauge group. As a consequence, the resulting three-dimensional model will be Lorentz-invariant. We assume that the four-dimensional spacetime has the topology $\mathcal{M}_4=\mathcal{M}_3\times\mathbb{S}^1$ where $\mathcal{M}_3$ is a three-dimensional spacetime, and $\mathbb{S}^1$ is space-like with coordinates $x^3$. In this way, we single out the third spatial component $\mu=3$. Let us now impose the conditions
\be\label{symmetry conditions}
\partial_3=0,\qquad\omega_3^{IJ}=0.
\ee
The first condition means that the fields do not depend on the third spatial direction $x^3$. The second one means that the parallel transport along  $\mathbb{S}^1$ is trivial. Therefore, the covariant derivative of the fields along the direction $\mu=3$ vanishes. A direct calculation shows that the four-dimensional Holst action reduces under the conditions (\ref{symmetry conditions}) to
\be\label{action form}
S_\text{red}=-\int_{\mathbb{S}^1}\de x^3\int_{\mathcal{M}_3}\de^3x\,\eps^{\mu\nu\rho}\left(\f{1}{2}\eps_{IJKL}e_3^Ie^J_\mu F^{KL}_{\nu\rho}+
\frac{1}{\gamma}\delta_{IJKL}e_3^Ie^J_\mu F^{KL}_{\nu\rho}\right),
\ee
where $\mu=0,1,2$ is now understood as a three-dimensional spacetime index, and $\de^3 x\,\eps^{\mu\nu\rho}$ is the local volume form on the three-dimensional
spacetime manifold. Apart from a global multiplicative factor that is not relevant at all, and provided that we set $x^I\equiv e_3^I$, we recover the three-dimensional action with Barbero-Immirzi parameter introduced in \cite{GN,GN2}, i.e.
\be\label{3D action}
S[e,x,\omega]=\int_{\mathcal{M}_3}\de^3x\,\eps^{\mu\nu\rho}\left(\frac{1}{2}\eps_{IJKL}x^Ie_\mu^JF_{\nu \rho}^{KL}+\frac{1}{\gamma}\delta_{IJKL}x^Ie_\mu^JF_{\nu \rho}^{KL}\right).
\ee
From now on, we will denote the three-dimensional spacetime manifold $\mathcal{M}_3$ simply by $\mathcal{M}$.

\subsection{Lagrangian analysis}

\noindent It is not immediately obvious that the action (\ref{3D action}) is equivalent to that of three-dimensional gravity, simply because its expression is rather different from the standard first order BF action. First of all, it seems that we have introduced an additional degree of freedom represented by the variable $x$, and secondly the internal gauge group is $\SL(2,\mathbb{C})$ instead of the usual gauge group $\SU(1,1)$ of Lorentzian three-dimensional gravity. Furthermore, the action now features a Barbero-Immirzi parameter\footnote{A Barbero-Immirzi-like parameter was previously introduced in \cite{Bonzon Livine} in the context of three-dimensional gravity, based on the existence of two independent bilinear invariant forms on the symmetry group of the Chern-Simons formulation. Unfortunately, this parameter does not feature the properties of its four-dimensional counterpart appearing in the Holst action. In particular, it does not disappear when one passes from the first to the second order formulation of the theory.}. Despite all these differences, it can be shown that the action (\ref{3D action}) represents a valid formulation of three-dimensional gravity \cite{GN,GN2}.

There are many ways to see that this is indeed the case. The easiest one consists in showing that the action (\ref{3D action}) reproduces the standard Einstein-Hilbert action when one goes from the first order to the second order formulation. This method does also show straightforwardly that the parameter $\gamma$ disappears exactly as it does in four dimensions, i.e. when one expressed the theory in the metric form. To make this statement concrete, it is convenient to decompose the connection $\omega$ into its self-dual and anti self-dual components $\omega^\pm$ according to the decomposition of $\sl(2,\mathbb{C})=\su(2)_\mathbb{C}\oplus\su(2)_\mathbb{C}$ into its self-dual and anti-self-dual complex subalgebras (see appendix \ref{appendix:algebra}). Then, the action (\ref{3D action}) can be expressed as a sum of two related BF action as follows:
\be\label{3d action as two BF}
S[e,x,\omega]=\left(1+\frac{1}{\gamma}\right)S[B^+,\omega^+]+\left(1-\frac{1}{\gamma}\right)S[B^-,\omega^-],
\ee
where $S[B^\pm,\omega^\pm]$ is the standard $\su(2)_\mathbb{C}$ BF action
\be
S[B^\pm,\omega^\pm]=\frac{1}{2}\int_\mathcal{M}\de^3x\,\eps^{\mu \nu \rho}\,\text{tr}\left(B_\mu^\pm,F_{\nu \rho}^\pm\right),
\ee
and
\be
B_\mu^{\pm i}=\pm\mathrm{i}(x\time e_\mu)^i+x^0e_\mu^i-x^ie^0_\mu
\ee
is calculated from the relations $B_\mu^{\pm i}=B^{IJ}_\mu T^{\pm i}_{IJ}$ and $B^{IJ}_\mu=\eps^{IJ}_{~~KL}x^Ke^L_\mu$. In this BF action, the trace $\tr$ denotes the normalized Killing form on $\su(2)_\mathbb{C}$. 

As usual, going from the first order to the second order formulation of gravity requires to solve for the components of the connection $\omega^\pm$ in terms of the $B$ variables. This can be done by solving the equations of motion obtained by varying the action with respect to the connection $\omega^\pm$, which are nothing but the torsion-free conditions
\be
T(B^\pm,\omega^\pm)=0.
\ee
If $\det(B^\pm)\neq0$, this torsion-free condition can be inverted to find the torsion-free spin connection $\omega(B)$. This latter, when plugged back into the original action (\ref{3d action as two BF}), leads to the sum of two second order Einstein-Hilbert actions,
\be\label{einstein-hilbert}
S_\text{EH}[g_{\mu\nu}^+,g_{\mu\nu}^-]=\f{1}{2}\left(1+\f{1}{\gamma}\right)\epsilon^+\int_\mathcal{M}\de^3x\sqrt{|g^+|}\,\mathcal{R}[g_{\mu\nu}^+]+\f{1}{2}\left(1-\f{1}{\gamma}\right)\epsilon^-\int_\mathcal{M}\de^3x\sqrt{|g^-|}\,\mathcal{R}[g_{\mu\nu}^-],
\ee
each being defined with respect to an Urbantke-like metric \cite{urbantke} $g_{\mu\nu}^\pm=B_\mu^\pm\cdot B_\nu^\pm$. In this expression, $\epsilon^\pm$ denotes the sign of $\det(B^\pm)$. It is straightforward to show that the signs $\epsilon^\pm$ are identical \cite{GN2}. To see that this is indeed the case, one can write the fields $B^\pm$ as follows:
\be\label{BdecompositionG}
B^{\pm i}_\mu=\pm\mathrm{i}\eps^i_{~jk}x^j e_\mu^k+x^0e^i_\mu-x^ie^0_\mu=\big(\pm\mathrm{i}x_0^{-1}\underline{x}+\openone\big)L_\mu^i,
\ee
with $L_\mu^i\equiv\eps^i_{~jk}B^{jk}_\mu/2=x^0e^i_\mu-x^ie^0_\mu$, and where we have introduced the three-dimensional matrix
\be
\underline{x}=
 \begin{pmatrix}
0 & -x_3 & x_2\\
x_3 & 0 & -x_1\\
-x_2 & x_1 & 0
 \end{pmatrix}
\ee
associated to $x$ such that $\underline{x}\alpha^i=\eps^i_{~jk}x^j\alpha^k$ for any $\alpha\in\mathbb{R}^3$. With this notation, we can compute the determinant
\be
\det(B^\pm)=\det\big(\pm\mathrm{i}x_0^{-1}\underline{x}+\openone\big)\det(L^i_\mu)=\big(1-x_0^{-2}(x_1^2+x_2^2+x_3^2)\big)\det(L^i_\mu),
\ee
where $L_\mu^i$ is considered as a $3\times3$ matrix. Therefore, we conclude that $\epsilon^+=\epsilon^-$, as announced above. Furthermore, a simple calculation shows that the two Urbantke metrics $g_{\mu\nu}^\pm$ are identical and given by
\be\label{urbantke metrics}
g_{\mu\nu}^\pm\equiv g_{\mu\nu}=(e_\mu\cdot e_\nu)\big(x^2-(x^0)^2\big)-x^2e^0_\mu e^0_\nu+x^0x\cdot(e^0_\mu e_\nu+e^0_\nu e_\mu),
\ee
with $x^2\equiv x^i x_i$.

Gathering these results on the Urbantke metrics and the sign factors $\epsilon^\pm$, we can conclude that, once the torsion-free condition is imposed, the action (\ref{3D action}) reduces to the standard Einstein-Hilbert action
\be
S_\text{EH}[g_{\mu\nu}]=\int_\mathcal{M}\de^3x\sqrt{|g|}\,\mathcal{R}[g_{\mu\nu}].
\ee
This shows that the theory that we are dealing with corresponds indeed to three-dimensional gravity, and that the Barbero-Immirzi parameter disappears once the torsion is vanishing. This is exactly what happens with the four-dimensional Holst action, which motivates the use of the three-dimensional model (\ref{3D action}) in order to test the role of the Barbero-Immirzi parameter.

\subsection{Lagrangian symmetries}

\noindent Before presenting the Hamiltonian analysis in details, let us finish the Lagrangian analysis with a study of the symmetries. This will be helpful in what follows. Obviously, the action (\ref{3D action}) is invariant under $\SL(2,\mathbb{C})$, and admits therefore the infinite-dimensional gauge group $\mathcal{G}\equiv C^\infty\big(\mathcal{M},\SL(2,\mathbb{C})\big)$ as a symmetry group. An element $\Lambda\in\mathcal{G}$ is an $\SL(2,\mathbb{C})$-valued function on the spacetime $\mathcal{M}$, which acts on the dynamical variables according to the transformation rules
\be\label{gauge}
e_\mu\longmapsto\Lambda\cdot e_\mu,\qquad
x\longmapsto\Lambda\cdot x,\qquad
\omega_\mu\longmapsto\text{Ad}_\Lambda(\omega_\mu)-\partial_\mu\Lambda\Lambda^{-1},
\ee
where $(\Lambda\cdot v)^I=\Lambda^{IJ}v_J$ denotes the fundamental action of $\Lambda$ on any four-dimensional vector $v$, and $\text{Ad}_\Lambda(\xi)=\Lambda\xi\Lambda^{-1}$ is the adjoint action of $\SL(2,\mathbb{C})$ on any Lie algebra element $\xi\in\sl(2,\mathbb{C})$.  

From the expression (\ref{3D action}) of the action as the integral of a three-form, it is immediate to see that the theory is also invariant under spacetime diffeomorphisms, as it should be for gravity. Infinitesimal diffeomorphisms are generated by vector fields $v=v^\mu\partial_\mu$ on $\mathcal{M}$, and their action on the dynamical variables is simply given by the following Lie derivatives:
\be\label{diffeos}
e\longmapsto\mathcal{L}_ve,\qquad
x\longmapsto\mathcal{L}_vx=v^\mu\partial_\mu x,\qquad
\omega\longmapsto\mathcal{L}_v\omega, 
\ee
where $\mathcal{L}_v\varphi=(v^\nu\partial_\nu\varphi_\mu+\varphi_\nu\partial_\mu v^\nu)\de x^\mu$ for any one-form $\varphi$.

The previous symmetries are expected from a theory of gravity formulated in first order variables. But a theory in three spacetime dimensions with only these symmetries would reduce to $\SL(2,\mathbb{C})$ BF theory, which is not what our model is. Thus, our Lagrangian should admit additional symmetries. This is indeed the case, and it is immediate to notice that the action (\ref{3D action}) is invariant under a rescaling symmetry and a translational symmetry. The former is generated by non-vanishing scalar fields $\alpha$ on $\mathcal{M}$ according to the transformation rules
\be\label{rescale}
e^I_\mu\longmapsto\alpha e^I_\mu,\qquad
x^I\longmapsto\frac{1}{\alpha}x^I.
\ee
The translational symmetry is generated by one-forms $\beta=\beta_\mu\de x^\mu$ according to
\be\label{extra trans}
e^I_\mu\longmapsto e^I_\mu+\beta_\mu x^I.
\ee
The presence of these two symmetries follow from the fact that the variables $x$ and $e$ appear in the action (\ref{3D action}) in the form $x^{[I}e^{J]}=(x^I e^J-x^J e^I)/2$. Note that they do not affect the connection $\omega$.

The transformations (\ref{gauge}), (\ref{diffeos}), (\ref{rescale}), and (\ref{extra trans}), encode all the symmetries of the action. We will make use of some of these invariance properties to simplify the canonical analysis in the following section. Furthermore, due to the $\SL(2,\mathbb{C})$ invariance, the sign of $x^2=x^Ix_I=x^I\eta_{IJ}x^J$ is an invariant of the theory, even if its value is not fixed because of the rescaling invariance. Thus, to define the theory, one has to fix this sign, and we choose it to be positive:
\be\label{sign condition}
x^2=x^I \eta_{IJ}x^J>0.
\ee
As we will see in the next section, this choice will make the time gauge accessible.

\section{The different gauge choices}
\label{sec:gauge choices}

\noindent In \cite{Sergey unpublished}, S. Alexandrov performed the Hamiltonian analysis of (\ref{3D action}) without assuming any gauge fixing. This analysis allows one to recover at the Hamiltonian level all the symmetries that were presented above. The formulation of the resulting phase space is unfortunately rather involved, and the present study does not require its full knowledge. For this reason, we will not present it here.

In \cite{GN}, the canonical analysis of (\ref{3D action}) was performed in two different gauges. These gauge choices have the advantage of simplifying the analysis drastically, and furthermore they reveal an intriguing interplay between the gauge fixing and the role of the Barbero-Immirzi parameter. With the first gauge choice of \cite{GN}, the initial gauge group $\SL(2,\mathbb{C})$ is broken into its (maximal non-compact) subgroup $\SU(1,1)$ (at least when the original action is chosen to be Lorentzian, otherwise one ends up with $\SU(2)$ instead), and the canonical analysis leads to a phase space with no $\gamma$-dependency. In this case, the Barbero-Immirzi parameter drops out of the theory already at the kinematical level. The second gauge choice of \cite{GN} is the three-dimensional analogue of the four-dimensional time gauge. It reduces $\SL(2,\mathbb{C})$ to its (maximal compact) subgroup $\SU(2)$. In this case, just like in four-dimensional loop gravity, the phase space has an explicit $\gamma$-dependency, both in the Poisson bracket between the connection and the densitized triad, and in the scalar constraint.

Before going any further, let us make a comment that will clarify a potential confusion between the role of the gauge groups. In \cite{GN}, the three-dimensional Holst action (\ref{3D action}) was chosen to be Euclidean, with gauge group $\SO(4)$. Because of this choice, the two gauge fixings described in the previous paragraph both lead to the subgroup $\SU(2)$. However, since in the present work we choose the gauge group of (\ref{3D action}) to be $\SL(2,\mathbb{C})$, the two gauge fixings lead to two different subgroups: $\SU(1,1)$ in the first case, and $\SU(2)$ in the time gauge.

This Section is devoted to reviewing the canonical analysis in these two gauges. The spacetime is assumed to have the topology $\mathcal{M}=\Sigma\times\mathbb{R}$ where $\Sigma$ is a space-like surface.

\subsection{The non-compact gauge: from the Lorentz group to SU(1,1)}
\label{sec:non-compact-gauge}

\noindent In this subsection, $\mu,\nu,\dots\in\{0,1,2\}$ are three-dimensional spacetime indices, $a,b,\dots\in\{1,2\}$ are spatial indices, $I,J,\dots\in\{0,1,2,3\}$ are internal $\SL(2,\mathbb{C})$ indices, and $i,j,\dots\in\{0,1,2\}$ are internal $\SU(1,1)$ indices. The indices $i,j,\dots$ are lowered and raised with the flat three-dimensional Minkowski metric $\eta_{ij}=\text{diag}(-1,+1,+1)$. We will use the cross-product notation $v\time w$ to denote the vector $z$ whose components are given by $z^i=\eps^{ijk}v_jw_k$, and $v\cdot w$ for the scalar product $v^iw_i=v^i\eta_{ij}w^j$.

The gauge group $\SL(2,\mathbb{C})$ is broken into the subgroup $\SU(1,1)$ by fixing in the action (\ref{3D action}) the field $x^I$ to the special value\footnote{Note the slight difference with \cite{GN}, where the gauge was chosen to be $x^I=(1,0,0,0)$.} $(0,0,0,1)$. This choice is compatible with the condition (\ref{sign condition}), and the rescaling symmetry (\ref{rescale}) can be used to fix the norm of $x$ to one for simplicity. The resulting $\SU(1,1)$ symmetry corresponds precisely to the isotropy group of $x$. Since this gauge choice singularizes the third internal space-like component, it is natural to decompose the connection $\omega^{IJ}$ into its $\su(1,1)$ components, denoted by $\omega^i$, and the complement denoted by $\omega^{(3)i}$. Therefore, we have
\be
\omega^i\equiv\frac{1}{2}\eps^i_{~jk}\omega^{jk},\qquad
\omega^{(3)i}\equiv\omega^{(I=3)i}.
\ee
The curvature tensor $F^{IJ}$ also decomposes into its $\su(1,1)$ components $F^i=\eps^i_{~jk}F^{jk}/2$, and the remaining part $F^{(3)i}$. Denoting by $F$ the vector with components $F^i$, and by $F^{(3)}$ the vector with components $F^{(3)i}$, we have the explicit expressions
\begin{subequations}
\ba
F_{\mu\nu}&=&\partial_\mu\omega_\nu-\partial_\nu\omega_\mu-\omega_\mu\time\omega_\nu-\omega^{(3)}_\mu\time\omega^{(3)}_\nu,\\
F^{(3)}_{\mu\nu}&=&\partial_\mu\omega^{(3)}_\nu-\partial_\nu\omega^{(3)}_\mu-\omega_\mu\time\omega^{(3)}_\nu+\omega_\nu\time\omega^{(3)}_\mu.
\ea
\end{subequations}
Using these notations, the action (\ref{3D action}) takes the following simple form:
\be
S=\int_\mathcal{M}\de^3x\,\eps^{\mu\nu\rho}e_\mu\cdot\left(F_{\nu\rho}+\frac{1}{\gamma}F^{(3)}_{\nu\rho}\right).
\ee
It is then immediate to see that neither $\omega^i_a$ nor $\omega^{(3)i}_a$ are the ``good'' dynamical connection variables of the theory. In fact, the canonical $\su(1,1)$ connection is given by the combination
\be\label{non-compact gauge connection A}
A^i_a\equiv-\left(\omega^i_a+\frac{1}{\gamma}\omega^{(3)i}_a\right),
\ee
which appears in the action through its curvature $\mathcal{F}$ as follows:
\be
S=-\int_\mathcal{M}\de^3x\,\eps^{\mu\nu\rho}e_\mu\cdot\left(\mathcal{F}_{\nu\rho}+\left(1+\frac{1}{\gamma^2}\right)\omega_\nu^{(3)}\time\omega_\rho^{(3)}\right).
\ee
In this way, $S$ has the form of an $\SU(1,1)$ BF action augmented with an extra term quadratic in $\omega^{(3)}$. Solving $\delta S/\delta\omega^{(3)i}=0$ implies that $\omega^{(3)}$ vanishes on-shell (assuming that $e$ is invertible), and therefore the theory becomes strictly equivalent to an $\SU(1,1)$ BF theory.

To perform the canonical analysis, we split the spacetime indices $\mu,\nu,\dots\in\{0,1,2\}$ into spatial indices $a,b,\dots\in\{1,2\}$ and the temporal direction denoted by $\mu=0$. With respect to this splitting, the action takes the following canonical form:
\be\label{canonical form}
S=\int_\mathbb{R}\de t\int_\Sigma\de^2x\left(E^a\cdot\partial_0A_a+A_0\cdot G+e_0\cdot H+2\left(1+\frac{1}{\gamma^2}\right)\omega^{(3)}_0\cdot\Phi\right),
\ee
where we have introduced the electric field $E^a=\eps^{ba}e_b$. The variables $A_0$, $e_0$ and $\omega^{(3)}_0$ are Lagrange multipliers enforcing the following primary constraints:
\be\label{constraints GHS}
G=\partial_aE^a+A_a\time E^a\simeq0,\qquad
H=\eps^{ab}\left(\mathcal{F}_{ab}+\left(1+\frac{1}{\gamma^2}\right)\omega_a^{(3)}\time\omega_b^{(3)}\right)\simeq0,\qquad
\Phi=E^a\time\omega^{(3)}_a\simeq0.
\ee
From the canonical form of the action, we can see that the only dynamical variables are the electric field $E^a$ and its canonically conjugated connection $A_a$. However, from the point of view of the canonical analysis, one has to consider $\omega_{a}^{(3)}$ as a dynamical variable as well, and introduce its conjugated momenta $\pi^a$ together with the constraints
\be\label{constraint pi}
\pi^a\simeq0
\ee
enforced by multipliers $\mu_a$. As a consequence, the symplectic structure is defined by the following Poisson brackets:
\be\label{variables in non compact}
\{E^a_i(x),A_b^j(y)\}=\delta^a_b\delta^j_i\delta^2(x-y)=\{\pi^a_i(x),\omega_b^{(3)j}(y)\},
\ee
and the time evolution $\partial_0\varphi$ of any field $\varphi$ it is defined from the total Hamiltonian
\be
H_\text{tot}=-\int_\Sigma\de^2x\left(A_0\cdot G+e_0\cdot H+2\left(1+\frac{1}{\gamma^2}\right)\omega^{(3)}_0\cdot\Phi+\mu_a\cdot\pi^a\right)
\ee
according to $\partial_0\phi=\{H_\text{tot},\phi\}$.

Among all the constraints (\ref{constraints GHS}) and (\ref{constraint pi}), only the last one, $\pi^a\simeq0$, implies secondary constraints. Imposing that its time evolution $\partial_0\pi^a_i=\{H_\text{tot},\pi^a_i\}$ be vanishing leads to the equations 
\be\label{equation P}
P^a\equiv\eps^{ab}\left(\omega_b^{(3)}\times e_0-\omega^{(3)}_0\time e_b\right)\simeq0.
\ee
These 6 equations involve Lagrange multipliers as well as dynamical variables, and as such they can be separated into two sets. The first set of equations fixes the values of Lagrange multipliers, and the second set is formed by secondary constraints. To extract these secondary constraints, it is convenient to combine the 6 equations (\ref{equation P}) with the 3 primary constraints $\Phi\simeq0$ given by (\ref{constraints GHS}). Indeed, these 9 equations can be written in the form
\be\label{S and P}
\eps^{\mu\nu\rho}e_\nu\time\omega_\rho^{(3)}\simeq0.
\ee
As a consequence, if $e$ is invertible (which is what we are assuming from the beginning), the original 9 equations (\ref{S and P}) are equivalent to the 9 equations $\omega^{(3)}_\mu \simeq 0$. It is then clear that the vanishing of $\omega^{(3)}_0$ is a fixation of Lagrange multipliers, whereas the remaining 6 equations $\omega^{(3)}_a\simeq0$ are (a mixture of primary and secondary) constraints. Moreover, these constraints together with (\ref{constraint pi}) clearly form a second class system, and can be solved strongly. Setting $\omega^{(3)}_a$ to zero in (\ref{canonical form}) shows that the Barbero-Immirzi parameter disappears completely, and we end up with the standard action of Lorentzian three-dimensional gravity. This closes the canonical analysis of the action (\ref{3D action}) in the the non-compact gauge.

This result is consistent with the observation that we made earlier at the Lagrangian level concerning the irrelevance of the Barbero-Immirzi parameter in the classical theory. As a consequence, $\gamma$ will play no role in the canonical quantum theory once we work in the non-compact gauge. This is already an interesting observation, since it seems to be in conflict with the situation in four dimensions where $\gamma$ plays a crucial role (at least) at the kinematical level. However, to make this conclusion stronger and more meaningful, we have to cast our three-dimensional model in a form that is closer to the four-dimensional Ashtekar-Barbero phase space, and then take this as the starting point for the quantization. This can be done by using the three-dimensional time gauge, as we will show in the next subsection.

\subsection{The time gauge: from the Lorentz group to SU(2)}
\label{sec:compact gauge}
	
\noindent The analysis of this subsection requires that we slightly change the notations for the internal indices. Now, we split the $\SL(2,\mathbb{C})$ indices into spatial indices $i,j,\dots\in\{1,2,3\}$ (which are interpreted as $\SU(2)$ indices) and the internal time direction $I=0$. The indices $i,j,\dots$ are lowered and raised with the flat three-dimensional Euclidean metric $\delta_{ij}=\text{diag}(+1,+1,+1)$. Once again, the cross-product  $z=v\time w$ will denote $z^i=\eps^i_{~jk}v ^jw ^k$, while the scalar product will be given by $v\cdot w=v^iw_i=v^i \delta_{ij}w^j$.

\subsubsection{The time gauge}

\noindent The gauge fixing that we refer to as the time gauge is defined by the requirement that $x^0=e_a^0=0$. This is clearly the three-dimensional analogue of the four-dimensional time gauge. This condition is compatible with (\ref{sign condition}) and breaks the gauge group $\SL(2,\mathbb{C})$ into its maximal compact subgroup $\SU(2)$. Since this gauge choice singles out the time component of $x$, it is now natural to decompose the connection $\omega^{IJ}$ into its spatial $\su(2)$ component $\omega^i$ and the complement denoted by $\omega^{(0)i}$. These components are given explicitly by
\be
\omega^i\equiv\frac{1}{2}\eps^i_{~jk}\omega^{jk},\qquad\omega^{(0)i}\equiv\omega^{(I=0)i}.
\ee
The curvature tensor $F^{IJ}$ does also decompose into its $\su(2)$ part, denoted with the same notation $F$ as in the previous subsection, and its temporal part $F^{(0)}$. Their respective components, $F^i=\eps^i_{~jk}F^{jk}/2$ and $F^{(0)i}$, are given by
\begin{subequations}
\ba
F_{\mu\nu}&=&\partial_\mu\omega_{\nu}-\partial_\nu\omega_{\mu}-\omega_{\mu}\time\omega_\nu+\omega^{(0)}_\mu\time\omega^{(0)}_\nu,\\
F^{(0)}_{\mu\nu}&=&\partial_\mu\omega^{(0)}_\nu-\partial_\nu\omega^{(0)}_\mu-\omega_\mu\time\omega^{(0)}_\nu-\omega_\nu\time\omega^{(0)}_\mu.
\ea
\end{subequations}

\subsubsection{Hamiltonian decomposition}

\noindent The action in the time gauge can be written in the following form:
\be
S=\int_\mathbb{R}\de t\int_\Sigma\de^2x\,( L_C+L_V+L_S),
\ee
where the ``canonical'', ``vectorial'', and ``scalar'' Lagrangian densities, respectively denoted by $L_C$, $L_V$, and $L_S$, are defined by
\begin{subequations}
\ba
L_C&=&2\eps^{ab}x\time e_a\cdot\left(F_{0b}^{(0)}-\frac{1}{2\gamma}F_{0b}\right),\\
L_V&=&\eps^{ab}x\time M\cdot\left(-F_{ab}^{(0)}+\frac{1}{\gamma}F_{ab}\right),\\
L_S&=&\eps^{ab}Nx\cdot\left(F_{ab}+\frac{1}{\gamma}F_{ab}^{(0)}\right).
\ea
\end{subequations}
The vectorial and the scalar Lagrangian densities are written respectively in terms of the vector $M^i=e_0^i$ and the Lapse function $N=e_0^0$. They encode the usual  vectorial and scalar constraints, as in the four-dimensional case. We will come back to their expression later on.

The canonical term $L_C$ tells us about the Poisson bracket and the Gauss constraint. Indeed, a straightforward calculation shows that, up to boundary terms that are assumed to be vanishing, $L_C$ takes the form
\be
L_C=\frac{2}{\gamma}\left(-E^a\cdot\partial_0A_a-\omega_0\cdot\big(\partial_aE^a+A_a\time E^a\big)
+\omega_0^{(0)}\cdot\big(\gamma\partial_aE^a-(\gamma\omega_a+\omega_a^{(0)})\time E^a\big)\right),
\ee
which implies that the canonical variables are given by
\be\label{A and E}
E^a\equiv\eps^{ab}e_b\time x,\qquad A_a\equiv\gamma\omega_a^{(0)}-\omega_a,
\ee
whereas $\omega_0$ and $\omega_0^{(0)}$ are Lagrange multipliers that enforce the primary constraints
\be\label{constraints G and Phi}
G=\partial_aE^a+A_a\time E^a\simeq0,\qquad
\Phi=\partial_aE^a-\omega_a\time E^a\simeq0.
\ee
Of course, $G\simeq0$ is the usual Gauss constraint, and we will show that $\Phi\simeq0$ is in fact a second class constraint. Note that $\Phi\simeq0$ is not equivalent to the constraint $\Phi'=\delta S/\delta\omega_0^{(0)}\simeq0$ that the Lagrange multipliers $\omega_0^{(0)i}$ impose, but it is instead a very simple linear combination of both $\Phi'\simeq0$ and $G\simeq0$.

The two remaining Lagrangian densities, $L_V$ and $L_S$, can also be expressed in term of the new connection variable $A$ (through its curvature $\mathcal{F}$), the associated electric field $E$ introduced in (\ref{A and E}), and the components $\omega$ of the initial $\sl(2,\mathbb{C})$ connection. To see that this is indeed the case, let us start by simplifying the expression of the vectorial density $L_V$. As in the four-dimensional case, it is useful to introduce the shift vector $N^a$ defined by $M^i=N^ae_a^i$, and then a straightforward calculation leads to the following expression for the vectorial Lagrangian density $L_V$:
\be
L_V=\frac{2}{\gamma}N^aE^b\cdot\left(-\mathcal{F}_{ab}+(\gamma^2+1)\omega_a^{(0)}\time\omega_b^{(0)}\right).
\ee
Due to the identity $E^a\cdot\omega_1^{(0)}\time\omega_2^{(0)}=-\gamma^{-1}\eps^{ab}\omega_b^{(0)}\cdot(G-\Phi)$, the Lagrangian $L_V$ takes the form
\be
L_V=-\frac{2}{\gamma}N^aE^b\cdot\mathcal{F}_{ab}+2\left(1+\frac{1}{\gamma^2}\right)N^a\omega_a^{(0)}\cdot(G-\Phi).
\ee
The scalar Lagrangian density is given by
\be
L_S=\frac{N}{\gamma}\eps^{ab}x\cdot\left(\mathcal{F}_{ab}+(1+\gamma^2)R_{ab}\right),
\ee
where $R_{ab}=\partial_a\omega_b-\partial_b\omega_a-\omega_a\times\omega_b$ is the curvature of the $\su(2)$ connection $\omega$. We see immediately that, as in four dimensions, $N$ and $N^a$ are Lagrange multipliers that enforce primary constraints. These take the following familiar form:
\be\label{constraints H}
H_a=\eps_{ab}E^b\cdot\mathcal{F}_{12},\qquad
H_0=x\cdot\left(\mathcal{F}_{12}+(1+\gamma^2)R_{12}\right). 
\ee
Notice that the variation of the action with respect to the shift vector $N^a$ does not lead directly to the previous expression for the vector constraint, but instead to a linear combination of $H_a$ with the primary constraints (\ref{constraints G and Phi}).

Finally, the analysis of the action in the time gauge shows that the theory can be formulated in terms of the variables $E^a_i$, $A_a^i$, $\omega_a^i$, $\omega_0^i$ and $\omega_0^{(0)i}$. Therefore, the initial connection components $\omega^{(0)}_a$ can be replaced by the three-dimensional version of the Ashtekar-Barbero connection, $A_a$, as it is the case in four-dimensions \cite{GN}. From the analysis of the canonical term, one can see that only $E$ and $A$ are a priori dynamical, whereas all the other variables have vanishing conjugate momenta. However, this does not mean that all these non-dynamical variables can be treated as genuine Lagrange multipliers. $\omega_0$ and $\omega_0^{(0)}$ can be treated as Lagrange multipliers, but $\omega_a$ has to be associated to a momentum $\pi^a$, as in the previous subsection. Because of this, the theory inherits new primary constraints enforcing the vanishing of $\pi^a$, i.e. 
\be\label{Pi constraint}
\pi^a\simeq0.
\ee
Up to a global factor of 2 in the action (that we can discard for the sake of simplicity), we end up with the following symplectic structure:
\be\label{conj variables}
\{E^a_i(x),A_b^j(y)\}=\gamma\delta^a_b\delta^j_i\delta^2(x-y),\qquad
\{\pi^a_i(x),\omega_b^{(0)j}(y)\}=\delta^a_b\delta^j_i\delta^2(x-y),
\ee
and the total Hamiltonian is given by
\be\label{Htot}
H_\text{tot}=\int_\Sigma\de^2x\left(\Lambda_0\cdot\Phi+\Omega_0\cdot G+N^aH_a-\frac{1}{\gamma}NH_0+\mu_a\pi^a\right),
\ee
and therefore appears as a linear combination of the primary constraints (\ref{constraints G and Phi}), (\ref{constraints H}) and (\ref{Pi constraint}). The Lagrange multipliers $\Omega_0$ and $\Lambda_0$ are also linear combinations of the Lagrange multipliers $\omega_0$, $\omega_0^{(0)}$, and $N^a$, and are given explicitly by
\be
\Omega_0=\frac{1}{\gamma}\omega_0-\frac{1}{\gamma^2}\omega_0^{(0)}+\left(1+\frac{1}{\gamma^2}\right)N^a\omega_a^{(0)},\qquad
\Lambda_0=\left(\frac{1}{\gamma^2}-1\right)\omega_0^{(0)}-\left(1+\frac{1}{\gamma^2}\right)N^a\omega_a^{(0)}.
\ee
We can now study the stability of the primary constraints.

\subsubsection{Analysis of the secondary constraints}

\noindent As in the four dimensional case, only the constraints $\pi^a\simeq0$ lead to secondary constraints. Their time evolution is given by the Poisson bracket with the total Hamiltonian (\ref{Htot}):
\ba
\partial_0\pi^a&=&\{H_\text{tot},\pi^a\}=\{\int_\Sigma\de^2x\left(\Lambda_0\cdot\Phi-\frac{1}{\gamma}NH_0\right),\pi^a\}\nonumber\\
&=&E^a\time\Lambda_0+\frac{1+\gamma^2}{\gamma}\eps^{ac}\big(x\partial_cN+N(\partial_cx-\omega_c\time x)\big).\label{do pi}
\ea
Requiring that $\partial_0\pi^a$ be vanishing gives rise to 9 equations involving Lagrange multipliers and dynamical variables. Among these equations, some determine the values of Lagrange multipliers, and others have to be interpreted as secondary constraints. To extract these secondary constraints, we first project the previous equations in the directions $E^b$ to obtain
\be
E^b\time E^a\cdot\Lambda_0+N(\partial_cx-\omega_c\time x)\cdot\eps^{ac}E^b\simeq0,
\ee
which eliminates the term proportional to derivatives of $N$ in (\ref{do pi}). Then, in order to eliminate the term proportional to $\Lambda_0$, we symmetrize the indices $a$ and $b$. If $N$ is non-vanishing, this leads to the 3 new constraints
\be\label{constraint Psi}
\Psi^{ab}=(\partial_cx-\omega_c\time x)\cdot\eps^{c(a}E^{b)}\simeq0.
\ee
This closes the Dirac analysis of the system, as one can show that there are no more constraints in the theory. In conclusion, the system is defined by the two pairs of conjugate variables (\ref{conj variables}) satisfying the constraints (\ref{constraints G and Phi}), (\ref{constraints H}), (\ref{Pi constraint}) and (\ref{constraint Psi}). Among these constraints, $\Phi$, $\pi$ and $\Psi$ form a complete set of $n_\text{S}=12$ second class constraints, and the other $n_\text{F}=6$ constraints are the first class generators of internal $\SU(2)$ gauge symmetry (i.e. the Gauss constraint) and of the spacetime diffeomorphisms (i.e. the vector and scalar constraints). Therefore, starting with $n_\text{NP}=2\times 12$ non-physical variables, we end up with $n_\text{P}=n_\text{NP}-2n_\text{F}-n_\text{S}=0$ local degrees of freedom, as it should be for three-dimensional gravity.

\subsubsection{Resolution of the second class constraints}

\noindent Before studying the quantization of the theory, one has to solve its second class constraints. There are two different but equivalent ways to deal with the second class constraints\footnote{One could also mention a third possibility, which is the so-called gauge unfixing procedure \cite{mitra-rajaraman}.}. The first one consists in computing the Dirac bracket. This makes the resolution of the second class constraints implicit in the sense that the Dirac bracket between any function on the phase space and a second class constraint strongly vanishes. This method, although being systematical, is sometimes very technical and far from the physical intuition. Furthermore, if the Dirac bracket is too complicated, the quantization can be very involved, as it is the case in Lorentz-covariant loop quantum gravity. If possible, one usually prefers to solve explicitly the second class constraints. This can be done in the present situation. The constraints $\pi^a=0$ are trivially resolved. The difficulty lies in the resolution of the 6 constraints $\Phi=0$ and $\Psi^{ab}=0$. These have to be understood as 6 equations for the 6 unknown components $\omega_a^i$ that we want to express in terms of $E$. It turns out to be easier to replace the 6 constraints $\Phi=0$ and $\Psi^{ab}=0$ by the following equivalent set of 6 equations:
\begin{subequations}
\begin{alignat}{2}
&E^1\cdot(\partial_2x-\omega_2\time x)=0,&\qquad\qquad
&E^2\cdot(\partial_1x-\omega_1\time x)=0,\label{eq 1}\\
&E^1\cdot(\partial_1x-\omega_1\time x)=0,&\qquad\qquad
&E^2\cdot(\partial_2x-\omega_2\time x)=0,\label{eq 2}\\
&E^1\cdot(\partial_aE^a-\omega_2\time E^2)=0,&\qquad\qquad
&E^2\cdot(\partial_aE^a-\omega_1\time E^1)=0,\label{eq 3}
\end{alignat}
\end{subequations}
whose equivalence with the original set can be shown through a simple calculation. To find the solution to these constraints, we use the fact that the family $(E^1,E^2,x)$ forms a basis of the internal space. This is the case because the original four-dimensional tetrad $e$ is supposed to be non-degenerate from the beginning. Furthermore, $x$ is orthogonal to the vectors $E^a$ by virtue of the definition (\ref{A and E}), and we have that
\be
E^1\time E^2=x(e_1\time e_2)\cdot x.
\ee
Since $x^i=e_3^i$ in the time gauge, the quantity $N(e_1\time e_2)\cdot x=N|e|$ is nothing but the determinant of the original tetrad. For this reason, we will use from now on the notation $|E|=(E^1\time E^2)\cdot x=x^2|e|$. 

The connection components $\omega_a$ can now be expended in the previous basis, and we can look for solutions of the form
\be\label{form of solutions}
\omega_a=\alpha_aE^1+\beta_aE^2+\zeta_ax,
\ee
where $\alpha_a$, $\beta_a$, and $\zeta_a$ are coefficients to be determined. Injecting this expression into equations (\ref{eq 1}) and (\ref{eq 2}), we immediately obtain the form of the coefficients $\alpha_a$ and $\beta_a$:
\be
\alpha_a=-\frac{1}{|E|}E^2\cdot\partial_ax,\qquad
\beta_a=\frac{1}{|E|}E^1\cdot\partial_ax.
\ee
Similarly, the coefficients $\zeta_a$ can be obtained by plugging (\ref{form of solutions}) into (\ref{eq 3}). This leads to
\be
\zeta_a=\frac{1}{|E|}\eps_{ab}E^b\cdot\partial_cE^c.
\ee
Gathering these results, we finally end up with the following solution for $\omega_a$:
\be
\omega_a=\frac{1}{|E|}\left(-(E^2\cdot\partial_ax)E^1+(E^1\cdot\partial_ax)E^2+\eps_{ab}(E^b\cdot\partial_cE^c)x\right),
\ee
which is the three-dimensional analogue of the four-dimensional Levi-Civita connection. The expression of $\omega_a$ is however much simpler in the present case, and it can even be further simplified to
\be\label{sol omega}
\omega_a=u\time\partial_au-(A_a\cdot u)u+\frac{1}{|E|}\eps_{ab}(E^b\cdot G)x,
\ee
where we have used the expression of the Gauss constraint, and introduced the unit normal $u=x/\sqrt{x^2}$ to the plane $(E^1,E^2)$ at each point. From equations (\ref{eq 1}) and (\ref{eq 2}), and the fact that $u$ is normalized, one can further show that the vector $u$ satisfies
\be
E^b\cdot(\partial_au-\omega_a\time u)=0,\qquad
u\cdot(\partial_au-\omega_a\time u)=0.
\ee
Since $(E^1,E^2,u)$ forms a basis of the three-dimensional internal space, we can finally conclude that $u$ satisfies the property
\be\label{property u}
\partial_au-\omega_a\time u=0.
\ee
This property will turn out to be very useful in the next section.

\subsubsection{Summary}

\noindent The resolution of the second class constraints ends the canonical analysis of the three-dimensional model in the time gauge. The phase space is now totally parametrized by the pair $(E^a_i(x),A_b^j(y))$, where $A$ is clearly the three-dimensional analogue of the $\su(2)$ Ashtekar-Barbero connection, and $E$ is its conjugate electric field. These variables satisfy the Poisson bracket
\be
\{E^a_i(x),A_b^j(y)\}=\gamma\delta^a_b\delta^j_i\delta^2(x-y),
\ee
and are subject to the constraints $G\simeq0$, $H_a\simeq0$, and $H_0\simeq0$. To cast the expression of the scalar constraint $H_0$ in a form similar to the one use in four-dimensional loop quantum gravity, let us modify the term $x\cdot R_{12}=\sqrt{x^2}\,u\cdot R_{12}$ in the following way:
\ba
u\cdot R_{12}&=&u\cdot(\partial_1\omega_2-\partial_2\omega_1-\omega_1\time\omega_2)\nonumber\\
&=&\partial_1(u\cdot\omega_2)-\partial_1u\cdot\omega_2-\partial_2(u\cdot\omega_1)+\partial_2u\cdot\omega_1-u\cdot\omega_1\time\omega_2\nonumber\\
&\simeq&-\partial_1(u\cdot A_2)+\partial_2(u\cdot A_1)+u\cdot\omega_1\time\omega_2\nonumber\\
&\simeq&-u\cdot(\partial_1A_2-\partial_2A_1)-\partial_1u\cdot A_2+\partial_2u\cdot A_1+u\cdot\omega_1\time\omega_2\nonumber\\
&\simeq&-u\cdot\mathcal{F}_{12}+u\cdot(A_1\time A_2+\omega_1\time A_2+A_1\time\omega_2+\omega_1\time\omega_2)\nonumber\\
&\simeq&-u\cdot\mathcal{F}_{12}+u\cdot K_1\time K_2.
\ea
In the second and fourth lines we have used the Leibniz rule, and in the third and fifth lines we have used the solution (\ref{sol omega}) and the property (\ref{property u}). Finally, in the last line, we have introduced the quantity
\be\label{Kformula}
K_a=A_a-(A_a\cdot u)u+u\time\partial_au=u\time(\partial_au+A_a\time u).
\ee
The weak equality $\simeq$ means equality up to terms proportional to the Gauss constraint $G$ or its derivatives $\partial_a G$. In this sense, the Hamiltonian constraint $H_0$ is therefore weakly equivalent to the following expression:
\be\label{modified H0}
u\cdot\big(\mathcal{F}_{12}-(1+\gamma^{-2})K_1\time K_2\big)\simeq0.
\ee
Since $K_a\simeq A_a+\omega_a$, the variable $\gamma^{-1}K_a$ can be seen as the three-dimensional analogue of the extrinsic curvature appearing in the four-dimensional Hamiltonian constraint. Because the vectors $K_a$ are orthogonal to $u$ by virtue of (\ref{Kformula}), the quantity $K_1\time K_2$ is clearly in the direction of $u$. As a consequence, one can view the vectorial constraints $H_a$ and the modified scalar constraint (\ref{modified H0}) as the components of the same three-dimensional constraint defined by
\be\label{F=0 in time gauge}
H=\mathcal{F}_{12}-(1+\gamma^{-2})K_1\time K_2\simeq0.
\ee
The fact that we can view $H_a$ and $H_0$ as the component of a same vector $H$ is a special property of this three-dimensional model that is evidently no longer true in four dimensions. As we will see, this is in some sense responsible for the fact that the model will be exactly solvable at the quantum level.

\subsection{On the role of the Barbero-Immirzi parameter}

\noindent Before studying the quantization of the theory, let us conclude this section about the classical analysis with a discussion on the role of the Barbero-Immirzi parameter. As already emphasized in \cite{GN} and reviewed in the previous subsections, the presence (or absence) of $\gamma$ in the description of the classical phase space seems to be closely related to the partial gauge fixing of the internal Lorentz group.

With the two gauge choices that we have just studied, it seems apparent that the three-dimensional Barbero-Immirzi parameter $\gamma$ ``knows something'' about the Lorentzian signature of the gauge group of the action (\ref{3D action}). Indeed, when using the time gauge and reducing the Lorentz group to $\SU(2)$, the Lorentzian signature is lost and, if $\gamma$ was not present, there would be no way of knowing that the original action that we started our analysis with was Lorentzian and not Euclidean\footnote{Just like in the four-dimensional Holst theory, once the gauge group (either $\SO(4)$ or $\SL(2,\mathbb{C})$) of the action is reduced to $\SU(2)$ by using the time gauge, the only remaining information at the level of the phase space about the gauge group of the non-gauge-fixed action is a relative sign in the scalar constraint.}. Therefore, everything happens as if $\gamma$ was keeping track of the fact that we started with a Lorentzian signature. By contrast, when $\SL(2,\mathbb{C})$ is reduced to $\SU(1,1)$ by using the gauge of subsection \ref{sec:non-compact-gauge}, the Lorentzian signature is still encoded in the gauge group after the gauge fixing, and $\gamma$ completely drops out of the theory because it becomes just superfluous.

Another important observation is that the connection variables $A$ (\ref{non-compact gauge connection A}) and (\ref{A and E}) that appear once we perform the two gauge choices have very different properties. In addition to their structure group being different because of the gauge fixing, their transformation behavior under diffeomorphisms differ. Indeed, the $\su(1,1)$ connection (\ref{non-compact gauge connection A}) transforms as a one-form under spacetime diffeomorphisms, whereas the $\su(2)$ connection (\ref{A and E}) transforms correctly only under spatial diffeomorphisms. Here again, the analogy between our model and the four-dimensional theory holds, and the anomalous transformation behavior of the $\su(2)$ connection is exactly analogous to the anomalous transformation behavior of the four-dimensional Ashtekar-Barbero connection \cite{alexandrov5,GN}. This comes from the well-known fact that the Ashtekar-Barbero connection is not the pullback of a spacetime connection \cite{samuel}.

What the three-dimensional model that we are studying here strongly suggests, is that $\gamma$ should be irrelevant at the quantum level. Indeed, we have seen that there exists a gauge in which the dynamical variable is an $\su(1,1)$ connection (which in addition transforms correctly under spacetime diffeomorphisms), and where $\gamma$ plays no role at all since it disappears already in the classical Hamiltonian theory. The quantization of this $\su(1,1)$ theory is far from being trivial, but it can be done for example using the combinatorial quantization scheme \cite{Alekseev:1994pa,Alekseev:1994au,BNR,Catherine}, and it is clear that $\gamma$ will play no role in this construction and not appear in the spectrum of any observable. This is a strong indication that $\gamma$ should not play any role at the quantum level even in the time gauge, if we require the $\SU(2)$ quantization to be anomaly-free. This would otherwise lead to anomalies, i.e. different quantum predictions in two different gauges, which is not physically acceptable. The two different gauge choices have to lead to equivalent physical predictions in the quantum theory. Therefore, either one can show that the imposition of all the quantum constraints in the $\SU(2)$ theory leads to the disappearance of $\gamma$ (which seems pretty unlikely), or the approach based on the $\SU(2)$ Ashtekar-Barbero connection has to be reconsidered and modified. Then, there could be two types of modifications. 1) One could think of abandoning the $\SU(2)$ connection at the classical level, and instead work with the self-dual connection and deal with the reality conditions. This is what we are going to do in the next section. 2) Alternatively, one can use $\gamma\in\mathbb{R}$ and the $\SU(2)$ formulation to start the quantization, but then an analytic continuation back to $\gamma=\pm\mathrm{i}$ has to be performed. We believe that this should be the case in the four-dimensional theory.

\section{Quantum theory}
\label{sec:quantum theory}

\noindent Since three-dimensional gravity admits only topological and no local degrees of freedom, for a long time if was thought to be too simple to be physically or mathematically interesting. The seminal work of Witten \cite{Witten 3d} based on its formulation as a Chern-Simons theory \cite{Achucarro-Townsend} showed that it was actually an exactly soluble system with incredibly rich underlying mathematical structures, and provided an unforeseen link with topological invariants \cite{Witten Jones}. This amazing result triggered an intense research activity around three-dimensional quantum gravity, which lead in particular to the introduction by Ponzano and Regge \cite{ponzano-regge} and later on Turaev and Viro \cite{turaev-viro}, of the first spin foam models. These models inspired later on, in four-dimensions, the attempts to represent the covariant dynamics of loop quantum gravity \cite{BC,EPR,EPRL,livine-speziale1,livine-speziale2,FK}, and in \cite{Noui:2004iy, Noui:2004iz} the link between three-dimensional loop quantum gravity and spin foam models was establish in the case of a vanishing cosmological constant (see \cite{AGN} for a more general review). This illustrates concretely the relevance of three-dimensional quantum gravity as a way to investigate the unknown aspects of the higher-dimensional theory. We show in this section that three-dimensional quantum gravity can also be used to investigate the role of the Barbero-Immirzi parameter in canonical loop quantum gravity.

This section is organized as follows. First we discuss the quantization of the three-dimensional model in the non-compact $\SU(1,1)$ gauge of subsection \ref{sec:non-compact-gauge}. We argue that the combinatorial quantization scheme can give a precise definition of the physical Hilbert space even if the gauge group is non-compact. By contrast, the loop quantization gives a clear definition of the kinematical Hilbert space but a more formal description of the physical Hilbert space. The rest of the section is devoted to the quantization in the $\SU(2)$ time gauge of subsection \ref{sec:compact gauge}. We adapt and apply the loop quantization by first turning the initial Ashtekar-Barbero connection into a complex (self-dual) connection, and then rewriting the associated reality conditions as a linear simplicity-like constraint. Finally, we show that the resolution of this constraint leads to the elimination of the Barbero-Immirzi parameter at the quantum level.

\subsection{Quantization in the non-compact gauge}

\noindent In this subsection, we recall a few facts about the quantization of the $G=\SU(1,1)$ BF theory that is obtained from the action (\ref{3D action}) in the non-compact gauge.

The total symmetry group $G_\text{tot}$ of a BF theory is bigger than the gauge group $G$, and is totally determined by the signature of the spacetime (or equivalently the ``signature'' of the gauge group $G$) and the sign of the cosmological constant $\Lambda$. In the case that we are interested in, $G=\SU(1,1)$ and $\Lambda=0$, and the total symmetry group  is the three-dimensional Poincar\'e group $G_\text{tot}=\ISU(1,1)$. In other words, $G$ is somehow augmented with the group of translations. The invariance under translations and the action of $G$ is equivalent (when the $B$ field satisfies invertibility properties) to the invariance under spacetime diffeomorphisms. The total symmetry group $G_\text{tot}$ has a clear geometrical interpretation as the isometry group of the three-dimensional Minkowski space $\mathbb{M}^3$, and any solution to the Einstein equations in the Lorentzian regime with vanishing cosmological constant is locally $\mathbb{M}^3$. In fact, such a BF theory is equivalent to a three-dimensional Chern-Simons theory whose gauge group is precisely $G_\text{tot}$. The Chern-Simons connection takes values in the Lie algebra $\su(1,1)\oplus\mathbb{R}^3$, and admits two components, an $\su(1,1)$ one and a translational one. The $\su(1,1)$ component is the original BF connection whereas the translational component is given essentially by the $B$ field (with the correct dimension).

The symmetry group $\ISU(1,1)$ is non-compact, and inherits the non-compactness of both $\SU(1,1)$ and the group of translations $\mathbb{R}^3$. This makes the quantization quite involved, and is the reason for which quantum BF theory was originally studied in the Euclidean case with a positive cosmological constant. Indeed, this is the only case in which the total symmetry group, $\SU(2)\times\SU(2)$, is compact. In this case, the path integral can be given a well-defined meaning, and gives (three-manifolds or knots) topological invariants. In the non-compact case the definition of the path integral is still an open problem. The most recent attempts to address this issue are based on analytic continuation methods to go from the compact case to the non-compact one \cite{Witten analytic}. To our knowledge, the Hamiltonian quantization offers a more efficient framework to study Chern-Simons theory with a non-compact group.

Among the different canonical quantizations methods for three-dimensional gravity, the loop and the combinatorial quantizations are certainly the most powerful ones. In fact, the two schemes are closely related as it was shown in \cite{Cath and I}. They are both based on a discretization of the spatial surface $\Sigma$, which is replaced by an oriented  graph $\Gamma$ sufficiently refined to resolve the topology of $\Sigma$. To simplify the discussion, we will assume that $\Sigma$ has no boundaries and does not contain any particles. Then, the graph $\Gamma$ is necessary closed, and contains $L$ links and $V$ vertices.

\subsubsection{The combinatorial quantization scheme}

\noindent The combinatorial approach consists in quantizing the theory in its Chern-Simons formulation. The dynamical variable is the $\isu(1,1)$ Chern-Simons connection, and to each oriented link $\ell$ of the graph $\Gamma$ is associated an element $U_\ell\in\ISU(1,1)$. After introducing a regularization scheme (based on the choice of a linear order at each vertex of $\Gamma$), the set of elements $U_\ell$ forms a quadratic Poisson algebra known as the Fock-Rosly Poisson bracket. The Fock-Rosly bracket involves classical $r$-matrices of $\isu(1,1)$, and its quantization naturally leads to the quantum double $\DSU(1,1)$ which plays a central role for the algebra of quantum operators. The precise definition of $\DSU(1,1)$ can be found for instance in \cite{Cath and I}, where it is shown that $\DSU(1,1)$ can be interpreted as a quantum deformation of the algebra of functions on the Poincar\'e group $\ISU(1,1)$. As a consequence, the combinatorial quantization clearly shows that, at the Planck scale, classical isometry groups are turned into quantum groups, and classical smooth (homogeneous) manifolds become non-commutative spaces. To make a very long story short, physical states are constructed from the representation theory of $\DSU(1,1)$.  The combinatorial quantization is a very powerful techniques that allows (at least in principle) to construct the physical Hilbert space of three-dimensional gravity for any Riemann surface $\Sigma$, even in the presence of point particles (see \cite{BNR} or \cite{Cath and I2} for instance). 

\subsubsection{The loop quantization}

\noindent The loop quantization is based on the BF formulation of three-dimensional gravity. In the continuous theory, the basic variables (\ref{variables in non compact}) are the $\su(1,1)$-valued connection $A$ and its conjugate variable $E$. Given a graph $\Gamma$, one introduces the holonomies $U_\ell\in\SU(1,1)$ along the links $\ell$ of $\Gamma$, and the ``fluxes'' $X_\ell$ of the electric field along edges dual to the links of $\Gamma$. These discretized variables $U_\ell$ and $X_\ell$ form the holonomy-flux Poisson algebra. The quantization promotes these classical variables to operators, the set of which forms a non-commutative algebra which can be represented, as usual in loop quantum gravity, on the Hilbert space
\be
\mathcal{H}_0(\Gamma)=\left(\mathcal{C}(\SU(1,1)^{\otimes L}),\de\mu(\Gamma)\right)
\ee
of continuous functions on the tensor product $\SU(1,1)^{\otimes L}$ endowed with the measure $\de\mu(\Gamma)$. At this stage, the measure is defined as the product of $L$ measures $\de\mu_0$ on $\SU(1,1)$. We notice immediately that the situation is more subtle than in the four-dimensional case because of the non-compactness of the group $\SU(1,1)$. Indeed, for the Hilbert space $\mathcal{H}_0(\Gamma)$ to be well-defined, one should restrict the space of continuous functions to the space of square integrable functions with respect to $\de\mu(\Gamma)$, i.e. $L^2\left((\SU(1,1)^{\otimes L},\de\mu(\Gamma)\right)$. However, any solution to the Gauss constraints is, by definition, invariant under the action of $\SU(1,1)$ at the vertices $v$ of $\Gamma$, and therefore cannot belong to the set of square integrable functions due to the infinite volume of $\SU(1,1)$. As a consequence, the construction of the kinematical Hilbert space requires a regularization process, which amounts to dividing out the volume of the gauge group. This has been studied and well-understood in \cite{Freidel Livine}. For this construction, it is useful to consider the simplest graph $\Gamma$ that resolves the topology of $\Sigma$. When $\Sigma$ is a Riemann surface (with no punctures and no boundaries) of genus $g$, the simplest graph $\Gamma$ consists in only one vertex $v$ and $L=2g$ loops starting and ending at $v$. Such a graph is called for obvious reasons a flower graph, and each loop is in one-to-one correspondence with a generator of the fundamental group $\Pi_1(\Sigma)$. Since the problem of defining the kinematical Hilbert space is a consequence of the invariance of kinematical states under the action of $\SU(1,1)$ at each vertex, this difficulty is considerably reduced by choosing $\Gamma$ to be a flower graph, and one can construct rigorously the kinematical Hilbert space
\be\label{kinematicalH}
\mathcal{H}_\text{kin}(\Gamma)=\left(\mathcal{\cal C}^\text{inv}(\SU(1,1)^{\otimes L}),\de\mu^\text{reg}(\Gamma)\right),
\ee
where ``inv'' stands for invariant and ``reg'' for regularized. One has
\be
f\in\mathcal{H}_\text{kin}(\Gamma)\quad\Longrightarrow\quad
f(U_1,\dots,U_L)=f(VU_1V^{-1},\dots,VU_LV^{-1}),\qquad
\int|f|^2\de\mu^\text{reg}(\Gamma)<\infty,
\ee
for $U_1,\dots,U_L$ and $V$ elements in $\SU(1,1)$. We refer the reader to \cite{Freidel Livine} for explicit details about this construction.

Once the Gauss constraint is imposed at the quantum level, the flatness condition has to be implemented. This was addressed in the Euclidean regime (where the gauge group is compact) in \cite{Noui:2004iy}. More precisely, it was shown that one can define a ``projector'' from the kinematical state space into the moduli space of flat $\SU(2)$ connections. This allows to construct rigorously the physical scalar product between kinematical states. The idea is very simple, and consists in replacing the measure $\de\mu_\text{kin}(\Gamma)$ on the kinematical Hilbert space associated to the graph $\Gamma$ by 
\be\label{physicalH}
\de\mu_\text{phys}(\Gamma)=\de\mu_\text{kin}(\Gamma)\prod_{f\in\Gamma}\delta\left(\overrightarrow{\prod_{\ell\subset f}}U_\ell\right),
\ee
where the first product runs over the set of faces $f$ in $\Gamma$ that can be represented by an ordered sequence $(U_1,\dots,U_n)$ of $n$ links, $\delta$ is the Dirac distribution on $\SU(2)$ and $U_\ell$ is the group element associated to the oriented link $\ell$. The physical scalar product can be shown (under certain hypothesis) to be well-defined, and to reproduce exactly the spin foam amplitudes of the Ponzano-Regge model. Even if one does not obtain generically (for any Riemann surface $\Sigma$) an explicit basis for the physical Hilbert space, one can concretely compute the physical scalar product between any two kinematical states. In principle, one could adapt this construction in order to define the physical scalar product in the non-compact case of $\SU(1,1)$ BF theory, and replace the measure on the kinematical Hilbert space (\ref{kinematicalH}) by a measure similar to (\ref{physicalH}) but with $\delta$ the Dirac distribution on $\SU(1,1)$ instead. Even if the presentation that we have done here is incomplete and formal, the technical details are not needed for the main purpose of the paper.

\subsection{Quantization in the time gauge}
	
\noindent We now study the quantization of the theory in the time gauge. There are essentially two ways of doing so. The first one is to mimic exactly four-dimensional loop quantum gravity, where one starts with the construction of the kinematical Hilbert space and then finds a regularization of the Hamiltonian constraint \`a la Thiemann in order to find the physical solutions. The second one relies on a reformulation of the classical phase space in a way that looks again like a BF theory. Let us start with a discussion about the first strategy. We use the same notations as in the previous subsection: $\Sigma$ is the spatial manifold, and $\Gamma$ a graph in $\Sigma$ with $L$ links and $V$ vertices.

In $\SU(2)$ loop quantum gravity, the construction of the kinematical Hilbert space leads to
\be\label{SU2 kinematical}
\mathcal{H}_\text{kin}(\Gamma)=\left(\mathcal{C}(\SU(2)^{\otimes L}),\de\mu(\Gamma)\right),
\ee
where $\mathcal{C}(G)$ denotes the space of continuous functions on the group $G$, and $\de\mu(\Gamma)$ is the usual Ashtekar-Lewandowski measure defined as a product of $L$ Haar measures $\de\mu_0$ on $\SU(2)$. Contrary to what happens in four dimensions where it is necessary to consider all possible graphs on $\Sigma$ (and to take a projective limit), here it is sufficient to fix only one graph (appropriately refined to resolve the topology of $\Sigma$) in order to define the kinematical Hilbert space. Since the gauge group is compact, the kinematical Hilbert space is well-defined, and $\mathcal{H}_\text{kin}(\Gamma)$ carries a unitary representation of the three-dimensional holonomy-flux algebra. The action of a flux operator $X_\ell$ on any kinematical state $\psi\in\mathcal{H}_\text{kin}(\Gamma)$ can be deduced immediately from the action on the representation matrices $\mathbf{D}^{(j)}(U_\ell)$ (which are the building blocks of the spin networks), where $\mathbf{D}^{(j)}:\SU(2)\rightarrow\mathbb{V}^{(j)}$ is the $\SU(2)$ spin-$j$ representation on the space $\mathbb{V}^{(j)}$ of dimension $d_j=2j+1$. This action is given by
\be\label{action of E}
X_\ell^i\triangleright\mathbf{D}^{(j)}(U_{\ell'})=-\mathrm{i}\gamma\lp\delta_{\ell,\ell'}\mathbf{D}^{(j)}(U_{\ell<c})J_i\mathbf{D}^{(j)}(U_{\ell>c}),
\ee
where $c$ denotes the intersection $\ell\cap\ell'$. The constants $\gamma$ and $\lp=\hbar G_\text{N}$ are the Barbero-Immirzi parameter and the three-dimensional Planck length. As in four dimensions, the spin network states diagonalize the three-dimensional analogue of the area operator, namely $\sqrt{X_\ell^2}$, whose eigenvalues are $\gamma\lp\sqrt{j_\ell(j_\ell+1)}$. Therefore, one arrives at the conclusion that in the time gauge the kinematical length operator has a discrete spectrum given by the Casimir operator of $\SU(2)$, and is furthermore proportional to the Barbero-Immirzi parameter, which can be interpreted as the fundamental length scale in Planck units. Just like in the four-dimensional case, one inherits a $\gamma$-dependency in the quantum theory, which as we will argue later on is completely artificial and an artifact of the gauge choice.

It is however legitimate to ask what happens if we try to push further the derivation of physical results based on this $\SU(2)$ formulation. For example, mimicking once again what is done in the four-dimensional theory, one could try to compute the entropy of a black hole, which in this three-dimensional model would correspond to a BTZ black hole. Using the notion of observables in the Turaev-Viro spin foam model, one can reproduce the calculation of \cite{BTZ-SF} and choose the fundamental length elements to be such that the perimeter $L$ of the black hole is given by
\be
L=8\pi\gamma\lp\sum_{\ell=1}^p\sqrt{j_\ell(j_\ell+1)},
\ee
where $p$ is the number of spin network links $\ell$ puncturing the horizon (we have here reintroduced the appropriate numerical factors). Then, the computation of the number of microstates leads at leading order to an entropy formula of the type
\be
S_\text{BH}=\f{L}{4\lp}\f{\gamma_0}{\gamma},
\ee
and one can proceed by fixing the value of the three-dimensional Barbero-Immirzi parameter to be $\gamma_0$, whose value can be computed explicitly. What is remarkable is that this value agrees with that derived in the four-dimensional case. This observation is a further indication that our three-dimensional model does indeed mimic exactly its four-dimensional counterpart, and that the behavior of the Barbero-Immirzi parameter is the same in both cases (once we use the time gauge and the $\SU(2)$ formulation).

Finally, once the kinematical structure is established, one should impose at the quantum level the three remaining constraints $H\simeq0$ (\ref{F=0 in time gauge}). These appear as the sum of two terms, $H=H_\text{E}-(1+\gamma^{-2})H_\text{L}$, where $H_\text{E}=\mathcal{F}_{12}$ and $H_\text{L}=K_1\time K_2$ are respectively called the Euclidean and the Lorentzian part of the constraints. In four dimensions, one has to consider separately the vector constraint and the scalar constraint, but we have seen that the peculiarity of three-dimensional gravity is that these can be treated as a single set. The set $H$ of constraints needs to be regularized in order to have a well-defined action on $\mathcal{H}_\text{kin}(\Gamma)$, and it is clear that the regularization of $H_\text{L}$ will lead to the same ambiguities that are present in four-dimensional canonical loop quantum gravity \cite{ale-amb,Bonzom:2011jv}. Since we know (from the quantization of three-dimensional gravity in the usual BF or spin foam setting) what the physical states should look like, one could potentially investigate these regularizations ambiguities of the Hamiltonian constraint and maybe try to clarify them. Although this would be a very interesting task that could have important consequences for the construction of the four-dimensional theory, we are going to follow instead the second strategy mentioned above, which consists in rewriting the $\SU(2)$ Ashtekar-Barbero phase space in the form of a BF theory.

\subsubsection{Equivalence with a complex BF theory}

\noindent Our aim is to reformulate the phase space of the three-dimensional theory in the time gauge in a way equivalent to a BF theory. More precisely, we are looking for a pair $(\mathbf{A}_a^i,\mathbf{E}^a_i)$ of canonical variables such that the constraints $G\simeq0$ (\ref{constraints G and Phi}) and $H\simeq0$ (\ref{F=0 in time gauge}) are equivalent to the constraints
\begin{eqnarray}\label{new constraints}
\mathbf{G}=\partial_a\mathbf{E}^a+\mathbf{A}_a\time\mathbf{E}^a\simeq0,\qquad
\mathbf{F}_{12}\simeq0, 
\end{eqnarray}
where $\mathbf{F}_{12}$ is the curvature of $\mathbf{A}$. We use the following ansatz for the expressions of the new variables in terms of the old ones:
\be
\mathbf{A}=A+\alpha L+\beta K,\qquad
\mathbf{E}=\zeta E+\xi u\time E,
\ee
where $L=\de u+A\time u$, $K=u\time L$ was introduced in (\ref{Kformula}), and $\alpha$, $\beta$, $\zeta$, and $\xi$ are constants that have to be fixed by the 
relations (\ref{new constraints}). In fact, this ansatz gives the most general expression for a connection $\mathbf{A}$ and an electric field $\mathbf{E}$ that transform correctly under $\SU(2)$ gauge transformations. This results from the fact that $u$, $L$ and $K$ transform as vectors under such gauge transformations.

It is useful to derive some properties of the quantities $L$ and $K$. A direct calculation shows that $L$ and $K$ satisfy the equations
\begin{subequations}
\ba
&&\partial_1L_2-\partial_2L_1+A_1\time L_2+L_1\time A_2=\mathcal{F}_{12}\time u,\\
&&\partial_1K_2-\partial_2K_1+A_1\time K_2+K_1\time A_2=u\time(\mathcal{F}_{12}\time u)+2L_1\time L_2.
\ea
\end{subequations}
Furthermore, we have that $L_1\time L_2 = K_1\time K_2$. Thus, the curvature $\mathbf{F}$ of $\mathbf{A}$ can be written in terms of $\mathcal{F}$ and $K$ as follows:
\be
\mathbf{F}_{12}=\mathcal{F}_{12}+(\alpha^2+\beta^2+2\beta)K_1\time K_2+\alpha\mathcal{F}_{12}\time u+
\beta(u\time\mathcal{F}_{12})\time u.
\ee
Due to the fact that $K_1\time K_2$ is in the direction of $u$, the curvature $\mathbf{F}$ takes the form
\be
\mathbf{F}_{12}=(1-\alpha\underline{u}-\beta\underline{u}^2)\big(\mathcal{F}_{12}+(\alpha^2+\beta^2+2\beta)K_1\time K_2\big),
\ee
where, for any vector $a$, $\underline{a}$ denotes the matrix that acts as $\underline{a}b_i=\eps_{i}^{~jk}a_jb_k$ on any vector $b$. As a consequence, since the three-dimensional matrix $(1-\alpha\underline{u}-\beta\underline{u}^2)$ is invertible, the constraints $H\simeq0$ are equivalent to $\mathbf{F}_{12}\simeq 0$ if and only if
\be
\alpha^2+(1+\beta)^2+\gamma^{-2}=0.
\ee
Before considering the fate of the Gauss constraint, we already see that the connection $\mathbf{A}$ must be complex when the Barbero-Immirzi parameter $\gamma$ is real. Indeed, the general solution of the previous equation is given by
\be
\alpha=z\sin\theta,\qquad
\beta=z\cos\theta-1,
\ee
with $z^2+\gamma^{-2}=0$, and where $\theta$ in an arbitrary angle. We will discuss the complexification in more detail later on.

Now, we compute the new Gauss constraint $\mathbf{G}$ in term of the original variables. A long but straightforward calculation shows that
\be
\mathbf{G}=\zeta G+\xi u\time G+(\zeta\beta-\xi\alpha) (G\cdot u)u+
(\xi+\alpha\zeta+\beta\xi)\big(\partial_au\time E^a+(A_a\cdot E^a)u\big),
\ee
which can be written as follows:
\be\label{capital G}
\mathbf{G}=MG+(\xi+\alpha\zeta+\beta\xi)\big(\partial_au\time E^a+(A_a\cdot E^a)u\big),
\ee
where $M$ is the matrix $M=\zeta(1+\beta)-\xi\alpha+\xi\underline{u}+(\zeta\beta-\xi\alpha)\underline{u}^2$. A necessary condition for $\mathbf{G} =0$ to be equivalent to $G=0$ is that the coefficient $(\xi+\alpha\zeta+\beta\xi)$ in front of the second term in (\ref{capital G}) be vanishing. This implies that $\xi=-\lambda\alpha$ and $\zeta=\lambda(1+\beta)$ with an arbitrary (but non-vanishing) coefficient $\lambda$ which in addition makes the matrix $M$ necessarily invertible. As a consequence, the general solution of the new constraints (\ref{new constraints}) is given by
\be\label{general sol}
\mathbf{A}=A+z\sin\theta\,L+(z\cos\theta-1)K,\qquad
\lambda^{-1}\mathbf{E}=z\cos\theta\,E+z\sin\theta\,(u\time E),
\ee
where $\theta$ is an arbitrary angle, $\lambda\neq0$, and $z^2+\gamma^{-2}=0$. Since $\lambda$ affects only the Poisson bracket between $\mathbf{A}$ and $\mathbf{E}$, we can set it to $\lambda=1$ for simplicity without loss of generality. 

At this point, there is a priori no reason for $\mathbf{A}$ and $\mathbf{E}$ to be canonically conjugated, and even $\mathbf{A}$ itself might be non-commutative. This would prevent the phase space of the theory in the time gauge from being equivalent to that of a BF theory. Fortunately, the previous expressions can be simplified considerably by noticing that all the solutions (\ref{general sol}) are in fact equivalent. More precisely, for any solution (\ref{general sol}), there exists a $\Lambda\in \SU(2)$ that sends this solution to the simple one corresponding to $\theta=0$. As a consequence, one can take $\theta=0$ without loss of generality, and this makes the study of the new variables much simpler. To see that this is indeed the case, let us compute how a solution (\ref{general sol}) transforms under the action of a rotation $\Lambda(n,\alpha)$ of angle $\alpha$ in the plane normal to $n$. Such an element is represented by the matrix
\be
\Lambda(n,\alpha)=\cos\left(\f{\alpha}{2}\right)+2\sin\left(\f{\alpha}{2}\right)J\cdot n
\ee
in the fundamental (two-dimensional) representation of $\SU(2)$, where $J_i$ are the $\su(2)$ generators satisfying the Lie algebra
\be
[J_i,J_j]=\eps_{ij}^{~~k}J_k.
\ee
If we identify any vector $a\in\mathbb{R}^3$ with an elements of $\su(2)$ according to the standard map $a\longmapsto a\cdot J$, the transformation laws for $\mathbf{A}\longmapsto\mathbf{A}^{\!\Lambda}$ and $\mathbf{E}\longmapsto\mathbf{E}^{\Lambda}$ under the action of $\Lambda$ are given by
\be\label{transform A and E}
\mathbf{A}^{\!\Lambda}=\Lambda^{-1}\mathbf{A}\Lambda+\Lambda^{-1}\de\Lambda,\qquad
\mathbf{E}^{\Lambda}=\Lambda^{-1}\mathbf{E}\Lambda.
\ee
To go further, we need to compute the adjoint action of $\SU(2)$ on its Lie algebra, and the differential form in the expression of $\mathbf{A}^\Lambda$:
\begin{subequations}
\ba
&&\text{Ad}_\Lambda J=\Lambda^{-1}J\Lambda=\cos\alpha\,J+\sin\alpha\,(n\time J)+2\sin^2\left(\f{\alpha}{2}\right)(n\cdot J)n,\\
&&\Lambda^{-1}\de\Lambda=\left(1+\sin^2\left(\f{\alpha}{2}\right)J\cdot n\right)\de\alpha+\sin\alpha\,(J\cdot\de n)-2\sin^2\left(\f{\alpha}{2}\right)J\cdot n\times\de n,
\ea
\end{subequations}
where we used the relation
\be
J_iJ_j=-\f{1}{4}\delta_{ij}+\f{1}{2}\eps_{ij}^{~~k} J_k
\ee
satisfied  by the 
$\su(2)$ generators in the fundamental representation. We can now compute the transformations (\ref{transform A and E}) to obtain
\begin{subequations}
\ba
\mathbf{A}^{\!\Lambda}&=&A+z\sin(\theta+\alpha)\,L+\big(z\cos(\theta+\alpha)-1\big)K,\\
\mathbf{E}^{\Lambda}&=&z\big(\cos(\theta+\alpha)\,E-\sin(\theta+\alpha)\,n\time E\big),
\ea
\end{subequations}
when $\alpha$ is assumed to be constant, i.e. $\de\alpha=0$. Taking $\alpha=-\theta$ simplifies the previous expressions and reduces the variables $\mathbf{A}^{\!\Lambda}$ and $\mathbf{E}^{\Lambda}$ to $\mathbf{A}^{\Lambda}$ and $\mathbf{E}^{\Lambda}$ given in (\ref{general sol}) where $\theta=0$. Finally, as announced, all the solutions of the type (\ref{general sol}) are equivalent. Therefore, we will now fix $\theta=0$, and use again the notation $\mathbf{A}$ and $\mathbf{E}$ to denote
\be\label{complex conn}
\mathbf{A}=A+(z-1)K,\qquad\mathbf{E}=zE.
\ee
As a conclusion, there is only one choice (up to $\SU(2)$ gauge transformations) of canonical variables that reduces the constraints obtained in the time gauge to BF-like constraints. However, it is immediate to notice that $\mathbf{E}$ and $\mathbf{A}$ are not canonically conjugated, and also that the components of the connection do not commute with respect to the Poisson bracket. This is a priori problematic since it makes the symplectic structure different from that of BF theory. Fortunately, there is a simple and very natural explanation for this fact. Instead of the connection $\mathbf{A}$, let us consider the connection
\be\label{asd}
\mathcal{A}_a=\mathbf{A}_a+\frac{z-1}{|E|}\eps_{ab}(E^b\cdot G)x=zA_a+(z-1)\omega_a,
\ee
which differs from $\mathbf{A}$ only by a term proportional to the Gauss constraint, and where $\omega^i_a$ is the solution of the second class constraints written in (\ref{sol omega}). Clearly, adding a term proportional to the Gauss constraint does not change anything to the previous analysis. Moreover, if we go back to the very first definition of $A^i_a=\gamma\omega^{(0)i}_a-\omega^i_a$ in terms of the boost $\omega^{(0)i}_a$ and rational $\omega^i_a$ components of the initial $\sl(2,\mathbb{C})$ connection, we see immediately that, depending on the sign of $\gamma z\in\{+,-\}$, the object
\be\label{shifted connection}
\mathcal{A}_a=zA_a+(z-1)\omega_a=\gamma z\omega^{(0)}_a-\omega_a=\pm\mathrm{i}\omega^{(0)}_a-\omega_a
\ee
is the (anti) self-dual component of the initial $\sl(2,\mathbb{C})$ connection. This comes from the fact that $z=\pm\mathrm{i}\gamma^{-1}$. In other words, reducing the phase space of the time gauge theory to that of a BF theory has mapped the initial $\su(2)$ Ashtekar-Barbero connection to the (anti) self-dual connection, as one could have anticipated. The property (\ref{shifted connection}) ensures that $\mathbf{E}$ and $\mathbf{A}$ (up to the Gauss constraint) satisfy the ``good'' canonical relations (a proof of this is given in appendix \ref{appendix:poisson}).

Since $z=\pm\mathrm{i}\gamma^{-1}$ is purely imaginary when the Barbero-Immirzi parameter $\gamma$ is real, $\mathbf{A}$ is complex and can be interpreted as an $\sl(2,\mathbb{C})$-valued connection. If we denote by $P_i \in \sl(2,\mathbb{C})$ the infinitesimal boost generators (see appendix \ref{appendix:algebra}) that satisfy the Lie algebra
\be
[J_i,P_j]=\eps_{ij}^{~~k}P_k,\qquad
[P_i,P_j]=-\eps_{ij}^{~~k}J_k,
\ee
and make explicit the Lie algebra generators that serve as a basis for the components of the connection, then the complex connection (\ref{complex conn}) can be identified with
\be\label{complex connection}
\mathbf{A}=A\cdot J+(z-1)K\cdot J=(A-K)\cdot J\pm\mathrm{i}\gamma^{-1}K\cdot J=(A-K)\cdot J\mp\gamma^{-1}K\cdot P
\ee
since $P=-\mathrm{i}J$ in the fundamental representation.

\subsubsection{Emergence of the linear simplicity-like condition}

\noindent Since the connection $\mathbf{A}$ is $\sl(2,\mathbb{C})$-valued, we cannot construct the quantum theory using the $\SU(2)$ kinematical Hilbert space (\ref{SU2 kinematical}), but instead we have to consider an extended space which is the space $\mathcal{H}_\text{kin}^\text{ext}(\Gamma)$ of $\SL(2,\mathbb{C})$ cylindrical functions. Without loss of generality, we can choose $\Gamma$ to be the flower graph that contains only one vertex. Elements of $\mathcal{H}_\text{kin}^\text{ext}(\Gamma)$ can be (formally) expanded into $\SL(2,\mathbb{C})$ spin networks, which consist of an assignment of irreducible representations of $\SL(2,\mathbb{C})$ to the links $\ell$, and of an intertwiner to the unique vertex $v$ of $\Gamma$. Any representation of $\SL(2,\mathbb{C})$ is labelled by a couple $(\chi_0,\chi_1)$ of complex numbers such that, for the principal series, $\chi_0=m+\mathrm{i}\rho$ and $\chi_1=-m+\mathrm{i}\rho$, with $\rho\in\mathbb{R}$ and $m\in\mathbb{N}$. Because of the non-compactness of the Lorentz group, it is difficult to treat the $\SL(2,\mathbb{C})$ invariance at the vertex of $\Gamma$ (i.e. the construction of $\SL(2,\mathbb{C})$ intertwiners) and to construct a positive-definite physical scalar product, but we will leave this technical difficulty aside. In fact, we are going to see that it is not necessary to consider the space of $\SL(2,\mathbb{C})$ states at all. The reason is that, as we will now show, the reality conditions satisfied by the complex connection ensure that it is $\su(1,1)$-valued. This $\SU(1,1)$ connection can then be taken as the starting point for the quantization.

Contrary to the $\SU(2)$ spin network states, the $\SL(2,\mathbb{C})$ ones contain a priori a non-trivial boost component. To understand this structure more precisely, let us decompose the connection $\mathbf{A}$ as follows:
\ba
\mathbf{A}&=&A\cdot J+(z-1)K\cdot J\nonumber\\
&=&A\cdot J+(\pm\mathrm{i}\gamma^{-1}-1)\big(u\times\de u+u\times(A\times u)\big)\cdot J\nonumber\\
&=&\left[-u\time\de u\cdot(J\pm\gamma^{-1}P)\right]+
\left[(A\cdot u)(u\cdot J)\mp\gamma^{-1}(A\time u)\cdot(P\time u)\right].\label{decomposition of A}
\ea
To obtain this expression, we have rewritten (\ref{complex conn}) with (\ref{Kformula}), used the relations $z=\pm\mathrm{i}\gamma^{-1}$ and $P_i=-\mathrm{i}J_i$, and made explicit the Lie algebra generators. It shows that $\mathbf{A}$ possesses two different parts, which are the two terms between the square brackets. Let us start by interpreting the second one. For this, we introduce the three $\sl(2,\mathbb{C})$ elements $\widetilde{J}_3=J\cdot u$ and $\widetilde{P}_\alpha=P\time u\cdot v_\alpha$, where the two vectors $v_\alpha^i\in\mathbb{R}^3$ ($\alpha=1,2$) are such that $v_\alpha\cdot v_\beta=\delta_{\alpha\beta}$ and $v_1\time v_2=u$. Therefore, $(v_1,v_2,u)$ forms an orthonormal basis of $\mathbb{R}^3$. The two vectors $v_\alpha$ are defined up to a rotation in the plane orthogonal to $u$, but this is not relevant for what follows. It is immediate to see that $(\widetilde{J}_3,\widetilde{P}_1,\widetilde{P}_2)$ forms the Lie algebra
\be
[\widetilde{P}_1,\widetilde{P}_2]=-\widetilde{J}_3,\qquad
[\widetilde{P}_2,\widetilde{J}_3]=\widetilde{P}_1,\qquad
[\widetilde{J}_3,\widetilde{P}_1]=\widetilde{P}_2,
\ee
and therefore generates an $\su(1,1)$ subalgebra of the initial $\sl(2,\mathbb{C})$ algebra. In the literature, the Lie algebra $\su(1,1)$ is usually defined as being generated by the three elements $(F_0,F_1,F_2)$ satisfying
\be
[F_1,F_2]=\mathrm{i}F_0,\qquad
[F_0,F_2]=\mathrm{i}F_1,\qquad
[F_0,F_1]=-\mathrm{i}F_2.
\ee
These generators are related to the previous ones through the map $(\widetilde{J}_3,\widetilde{P}_1,\widetilde{P}_2)\longmapsto\mathrm{i}(F_0,F_1,F_2)$.

It is now easy to see that the second term between square brackets in (\ref{decomposition of A}) defines an $\su(1,1)$-valued one-form which has to be interpreted as an $\su(1,1)$ connection. This condition is necessary in order to avoid anomalies, i.e. different quantum theories in two different gauges. Indeed, we saw in the previous section that there is a gauge in which the original three-dimensional Holst action takes the form of an $\SU(1,1)$ BF theory, and therefore it is natural to recover an $\su(1,1)$ connection even if we work in the time gauge. However, for this to be true, the first term in the expression (\ref{decomposition of A}) for the $\sl(2,\mathbb{C})$ connection must vanish, i.e. $(J\pm\gamma^{-1}P)\time u=0$. This relation means that the two (independent) components of $J\pm\gamma^{-1}P$ that are orthogonal to $u$ are constrained to vanish, while the third component is a priori unconstrained. In fact, this third component has to be constrained as well for the sake of consistency. To understand this point, the constraints $(J\pm\gamma^{-1}P)\time u=0$ have to be interpreted as reality conditions that select a real form in $\SL(2,\mathbb{C})$. There are only two possible ways of doing so, which correspond to selecting the subgroup $\SU(2)$ or the subgroup $\SU(1,1)$. These solutions are associated respectively with the linear simplicity constraints
\begin{subequations}\label{simplicity modified}
\ba
\SU(2)\quad:\quad&&\left\{(J\pm\gamma^{-1}P)\time u=0\,; (J\pm\gamma^{-1}P)\cdot u=0\right\},\label{simplicity modified compact}\\
\SU(1,1)\quad:\quad&&\left\{(J\pm\gamma^{-1}P)\time u=0\,; (P\mp\gamma^{-1}J)\cdot u=0\right\},\label{simplicity modified ncompact}
\ea
\end{subequations}
which represent the only two possible extensions of $(J\pm\gamma^{-1}P)\time u=0$. Interestingly, the constraints (\ref{simplicity modified compact}) can be written as
\be\label{su(2) simplicity}
J_i\pm\gamma^{-1}P_i=0,\qquad i\in\{1,2,3\},
\ee
which corresponds to the linear simplicity constraints of the EPRL and FK${}_\gamma$ spin foam models. On the other hand, the constraints (\ref{simplicity modified ncompact}) can be written as
\be
J_1\pm\gamma^{-1}P_1=0,\qquad J_2\pm\gamma^{-1}P_2,\qquad P_3\mp\gamma^{-1}J_3=0,
\ee
or equivalently in the more compact form
\be\label{su(1,1) simplicity}
G_i\pm\gamma^{-1}F_i=0,\qquad i\in\{0,1,2\}.
\ee
Here, $F_i$ are the three generators of $\su(1,1)$ introduced in (\ref{F generators}), and the elements $G_i$ span the complement of $\su(1,1)$ in $\sl(2,\mathbb{C})$, and are defined by $(G_0,G_1,G_2)=\mathrm{i}(P_3,-J_1,-J_2)$. This constraint was introduced in \cite{conrady} to define spin foam models for general Lorentzian four-geometries. One can check that these elements satisfy the commutation relations
\be\label{FG commutation relations}
[F_i,F_j]=\mathrm{i}c_{ij}^{~~k}F_k,\qquad[F_i,G_j]=\mathrm{i}c_{ij}^{~~k}G_k,\qquad i\in\{0,1,2\},
\ee
where the structure constants $c_{ij}^{~~k}$ are given in (\ref{su11 structure c}). These commutation relations tell us that (\ref{su(1,1) simplicity}) has the exact same structure as (\ref{su(2) simplicity}), in the sense that both sets of simplicity constraints relate two vectors that transform in a similar way under the action of a subgroup of $\SL(2,\mathbb{C})$ (i.e. either $\SU(2)$ or $\SU(1,1)$), one of these vectors being the generator of this given subgroup. In summary, we see that the constraints (\ref{simplicity modified}) select a subalgebra of $\sl(2,\mathbb{C})$.
\begin{itemize}
\item[1.] The constraints $J_i\pm\gamma^{-1}P_i=0$ consist in modifying the action of the infinitesimal boosts $P_i$ while keeping the action of the infinitesimal rotations $J_i$ unchanged in such a way that the relation (\ref{su(2) simplicity}) is satisfied, i.e. by setting $P_i=\mp\gamma J_i$. This is exactly what is done in the construction of spin foam models. By doing this, the action of the boosts is somehow compactified, and if one replaces $P_i$ by $\mp\gamma J_i$ in the expression (\ref{decomposition of A}), the $\sl(2,\mathbb{C})$ connection reduces to $\mathbf{A}=A\cdot J$. This solution is the initial $\su(2)$-valued Ashtekar-Barbero connection, and therefore it cannot lead to an $\su(1,1)$ connection.
In this sense, the first sector of solutions to the simplicity-like constraints (\ref{simplicity modified}) selects the maximal compact subalgebra $\su(2)$ of $\sl(2,\mathbb{C})$.
\item[2.] The constraints $G_i\pm\gamma^{-1}F_i=0$ consist in modifying the action of the infinitesimal generators $G_i$, while keeping unchanged the action of the infinitesimal $\su(1,1)$ generators $F_i$. In this sense, the rotations are ``decompactified'', and the complex connection (\ref{decomposition of A}) reduces to the non-compact element $\mathbf{A}=\mp\gamma^{-1}A\cdot P$. This object does not a priori define an $\su(1,1)$ connection since the $P$'s do not form a Lie subalgebra of $\sl(2,\mathbb{C})$. However, this is only an apparent problem. Indeed, let us recall that our extension from $\su(2)$ to $\sl(2,\mathbb{C})$ has been done in the two-dimensional representation. In this representation, the generators $P_i$ and $J_i$ are related by a global factor of $\mathrm{i}$, i.e. $P_i=-\mathrm{i}J_i$. As a consequence, in the fundamental representation, the connection satisfying the constraint (\ref{su(1,1) simplicity}) can be written in the form
\be\label{second solution}
\mathbf{A}=\mp\gamma^{-1}A\cdot P=\pm\mathrm{i}\gamma^{-1}A\cdot J=\mp\mathrm{i}\gamma^{-1}\left(A^1F_1+A^2F_2-\mathrm{i}A^3F_0\right),
\ee
where we have identified the $\su(1,1)$ generators $(F_0,F_1,F_2)$ with $-\mathrm{i}(J_3,P_1,P_2)=-(\mathrm{i}J_3,J_1,J_2)$. It is therefore clear that the second solution selects the non-compact $\su(1,1)$ connection in the initial Lorentz algebra.
\end{itemize}

We have presented here what seems to be the only two consistent ways of interpreting the constraints (\ref{simplicity modified}). Indeed, these constraints should select a three-dimensional subalgebra of $\sl(2,\mathbb{C})$ in order for the resulting connection to be well-defined. The first constraint, (\ref{simplicity modified compact}), selects the compact subalgebra and corresponds to the choice made in the EPRL and FK${}_\gamma$ spin foam models. In the context of our analysis, this constraint is not physically relevant since we expect the resulting connection to be valued in $\su(1,1)$ and not in the Lie algebra of a compact group. By contrast, the second constraint, (\ref{simplicity modified ncompact}), looks much more appealing since it leads to an $\su(1,1)$-valued connection and also to the disappearance of the Barbero-Immirzi parameter in the spectrum of the geometrical operators due to the overall factor of $\gamma^{-1}$ in $\mathbf{A}$ (\ref{second solution}). Thus, in the time gauge, we have been able to turn the initial Ashtekar-Barbero connection to an $\su(1,1)$ connection by going through a complexification and then imposing the reality conditions. This results is totally consistent with the results obtained in the non-compact gauge of section \ref{sec:non-compact-gauge}, where the Hamiltonian formulation was shown to be that of an $\SU(1,1)$ BF theory. Since the equivalence between the $\SU(2)$ theory in the time gauge and a BF theory is established in this three-dimensional model, the full quantization (i.e. the imposition of the quantum flatness constraint) can in principle be performed and does not pose any conceptual problems (even if it can be mathematically involved).

\subsubsection{Action of the flux operator}

\noindent Let us finish this section with a  quick discussion on the disappearance of the Barbero-Immirzi parameter from the spectra of the geometrical operators. We refer the reader to the companion paper \cite{BGNY} for more details. Once the constraints (\ref{su(1,1) simplicity}) are selected, we have the $\su(1,1)$ connection (\ref{second solution}) and the spin network states are colored with unitary irreducible representations of $\SU(1,1)$ that we label by $s$. We know that the action of the flux operator $X_\ell^i$ defined by the triad $E^a_i$ is given by $-\mathrm{i}\hbar\delta/\delta A^i_a$. Moreover, since $\lb E^a_i,A^j_b\rb=\gamma\delta^a_b\delta^j_i$, we have the Poisson bracket $\lb E^a_i,\mathbf{A}^j_b\rb=\pm\mathrm{i}\delta^a_b\delta^j_i$, and we can compute the action of the flux on the holonomy of the connection (\ref{second solution}) to find
\be\label{action of E}
X_\ell^i\triangleright\mathbf{D}^{(s)}(U_{\ell'})=\pm\lp\delta_{\ell,\ell'}\mathbf{D}^{(s)}(U_{\ell<c})J_i\mathbf{D}^{(s)}(U_{\ell>c}),
\ee
where $c$ denotes the intersection $\ell\cap\ell'$. An equivalent point of view would be to see the flux operator as acting on the holonomy of the shifted (or self-dual) connection $\mathcal{A}$ defined in (\ref{shifted connection}), whose Poisson bracket with $E$ is given by $\pm\mathrm{i}$. This action is also independent of $\gamma$. Therefore, we see that it is equivalent to consider the self-dual theory with an imaginary Barbero-Immirzi parameter or the complex theory defined with (\ref{second solution}) and $\gamma\in\mathbb{R}$, since in this case $\gamma$ disappears due to the redefinition of the appropriate variables. Beyond this observation about the role of $\gamma$, it is even more interesting to see that it is possible to obtain a positive-definite length spectrum.

Indeed, the ``gauge invariant'' quadratic operator $X_\ell^2$ is diagonalized by the spin networks and its eigenvalues are given by
\be
X_\ell^2\triangleright\mathbf{D}^{(s)}(U_{\ell'})=\lp^2\delta_{\ell,\ell'}Q^{(s)}\mathbf{D}^{(s)}(U_{\ell}),
\ee
where $Q^{(s)}$ denotes the evaluation of the $\SU(1,1)$ Casimir operator $Q=F_1^2+F_2^2-F_0^2$ in the representation labelled by $s$. Let us recall that there are two families of unitary irreducible representations of $\su(1,1)$, the continuous series (non-exceptional and exceptional classes) and the discrete series (positive and negative). $Q$ takes negative values for the later and positive values for the former. Therefore, if one requires that $X_\ell^2$ be a positive-definite operator, only the continuous series is admissible. Since $X_\ell^2$ is the building block of the length operator, at the physical level this operator will necessarily have a continuous spectrum. This is to be contrasted with the a priori prediction that could have been made at the kinematical level if we had stopped the analysis of the $\SU(2)$ theory in the time gauge at the end of section \ref{sec:gauge choices}, i.e. before recasting the Hamiltonian constraint as a flatness constraint for the complex connection $\mathbf{A}$, and before arriving at the linear constraint that selects the $\su(1,1)$ connection. Indeed, if we had stayed at the superficial level of the $\SU(2)$ kinematics, we would have derived a discrete length spectrum proportional to $\gamma$. The observation that working with the $\SU(1,1)$ representations leads to a continuous spectrum independent of $\gamma$ is completely consistent with the fact that we are describing Lorentzian three-dimensional gravity \cite{freidel-livine-rovelli}.

\section{Conclusion and perspectives}

\noindent In this paper we have studied the role of the Barbero-Immirzi parameter and the choice of connection in the construction of a symmetry reduced version of loop quantum gravity. This symmetry reduction consists in imposing invariance along a given spatial direction, which reduces the original four-dimensional Holst action to an action for three-dimensional gravity with a Barbero-Immirzi parameter. This action was originally introduced and analyzed in \cite{GN2} in its Plebanski form, and further studied in \cite{GN} in Euclidean signature and for two specific gauge choices. In the Lorentzian theory, these two gauge choices, which we have shown to be consistent with the dynamics of three-dimensional gravity, have drastically different interpretations. The first one, studied in section \ref{sec:non-compact-gauge}, reduces the action to that of $\SU(1,1)$ BF theory, and leads to a Hamiltonian formulation without any dependency on the Barbero-Immirzi parameter. The second one, studied in section \ref{sec:compact gauge} and which we refer to as the time gauge, leads just like in the four-dimensional case to an $\SU(2)$ theory written in terms of the Ashtekar-Barbero connection for $\gamma\in\mathbb{R}$ and admitting the same type of first class Gauss, scalar and vector constraints. Since three-dimensional gravity is an exactly soluble (classical and quantum) system, we have argued that this model can serve as a test bed to understand the relevance of the Barbero-Immirzi parameter in the dynamics of quantum gravity.

We have seen in this three-dimensional model that it is possible to rewrite the scalar and vector constraints of the $\SU(2)$ theory in the time gauge in the form of a unique flatness constraint for a complex connection $\mathbf{A}$, and that this latter is closely related to the complex (anti) self-dual Ashtekar-Barbero connection. However, nowhere did we set by hand the Barbero-Immirzi parameter to the value $\gamma=\pm\mathrm{i}$. Then, we have argued that in order for the quantum theory to be consistent with the quantization of Lorentzian three-dimensional gravity (i.e. $\SU(1,1)$ BF theory), the complex connection $\mathbf{A}$ had to be $\su(1,1)$-valued, a requirement that is met only if the generators of $\sl(2,\mathbb{C})$ satisfy the constraints $J\pm\gamma^{-1}P=0$ or $G\pm\gamma^{-1}F=0$. These constraints are nothing but the linear simplicity constraints used in the construction of four-dimensional spin foam models, and their (individual) role is to restrict the representations of the Lorentz group in a way that is compatible with the dynamics of quantum general relativity. More specifically, we have seen that among these two constraints, only $G\pm\gamma^{-1}F=0$ is consistent with the physical content of the theory. Indeed, it is this constraint that reduces the $\sl(2,\mathbb{C})$ connection to an $\su(1,1)$ connection, while the other one gives back an $\su(2)$ connection. Working with the connection (\ref{second solution}) then leads to continuous spectra for the kinematical (and a priori physical) geometrical operators and to the disappearance of the Barbero-Immirzi parameter.

These new and surprising observations raise a lot of questions, the most important one certainly being that of their implication for the four-dimensional theory and both its canonical and spin foam quantizations. At first sight, it may seem that we are running into circles. Indeed, it is known that in four-dimensional canonical gravity one can make the choice $\gamma=\pm\mathrm{i}$, which evidently gets rid of the Barbero-Immirzi ambiguity in the theory and simplifies the Hamiltonian constraint, but at the expense of introducing the reality conditions which we do not know how to implement at the quantum level. However, what our three-dimensional model has shown is that the reality conditions can in some sense be traded for the linear simplicity-like constraints $G\pm\gamma^{-1}F=0$. These can indeed be thought of as reality conditions since they constrain the components of the complex connection $\mathbf{A}$ in such a way that the resulting connection is $\su(1,1)$-valued (which, as we have argued, is a physically-consistent requirement since we are describing Lorentzian three-dimensional gravity). Moreover, we have seen that in order to obtain the $\su(1,1)$ connection, the simplicity constraint has to be interpreted in a different way from what is done in the EPRL and FK spin foam models, i.e. by selecting a non-compact subgroup of $\SL(2,\mathbb{C)}$ and not a compact one. Finally, we have pointed out that this construction leads to a positive-definite length spectrum, which translates the reality condition on the metric.

Without the new ingredient of our construction, which consists in sending the three-dimensional Ashtekar-Barbero phase space back to that of $\SU(1,1)$ BF theory, we would have derived a ``wrong" kinematical structure for three-dimensional gravity. Indeed, if we had worked with the $\SU(2)$ theory we would have obtained a discrete length spectrum proportional to $\gamma$. Alternatively, if we had naively chosen $\gamma=\pm\mathrm{i}$ in order to get rid of the Barbero-Immirzi ambiguity and simplify the Hamiltonian constraint, we would have constructed the kinematical structure with $\sl(2,\mathbb{C})$ spin network states and obtained an incorrect minus sign in the length spectrum (unless the representations entering the Casimir operator are interpreted differently). The key point is therefore the derivation of the simplicity-like constraint which selects the $\su(1,1)$ subalgebra as the kinematical arena on top of which to construct the physical Hilbert space.

Let us emphasize once again that the constraint (\ref{su(1,1) simplicity}) has been derived and imposed at the classical level, which is the reason why the kinematical states that we consider are $\SU(1,1)$ spin networks. Alternatively, one could think of constructing the kinematical states with the full $\sl(2,\mathbb{C})$ connection (\ref{decomposition of A}), and then imposing the simplicity constraint in the quantum theory as it is done in the construction of spin foam models. However, following this second approach one would run into the problem that the spin foam imposition of the constraints constrains the holonomies rather than the connection itself \cite{Alexandrov:2010pg}. The relationship between these two impositions of the constraints (constrain and quantize versus quantize and constrain) should be investigated further in this three-dimensional model, together with the comparison between the path integral and canonical quantizations. Additionally, it is important to point out that we did not try to solve the Ashtekar-Barbero Hamiltonian constraint (\ref{F=0 in time gauge}) in order to find the physical states and to study the fate of the Barbero-Immirzi parameter. Instead, we have argued that already at the classical level there is a different choice of connection to be made, and that this choice naturally comes with a set of simplicity-like constraints that can be imposed in a natural way.

As far as the full four-dimensional theory is concerned, we now have to think about the implementation of this three-dimensional construction in both the canonical theory and the spin foam models. It is quite likely that in the canonical theory there will be analogous simplicity-like conditions which restrict the type of representations that have to be considered (in fact this has already been observed in \cite{wieland}). The situation might be clearer in spin foam models, since their construction relies mainly on properties of the internal symmetry group and not that much on the symmetries of the spacetime, and our three-dimensional model has been constructed without affecting the internal symmetry group. Of course, the internal symmetry group has been affected by our gauge choice, but this is exactly what happens in spin foam models, where the simplicity constraints on the $B$ field induce relations between the $\SL(2,\mathbb{C})$ representations, which in turn define the $\SU(2)$ Ashtekar-Barbero connection starting from the initial Lorentz spin connection. It might very well be that when implementing the linear simplicity constraint with $\gamma=\pm\mathrm{i}$, one has to understand the resulting self-dual $\SL(2,\mathbb{C})$ representations rather as representations in the continuous series of $\SU(1,1)$.

This idea that the original complex Ashtekar variables may play an important role in quantum gravity has been recently revived in the context of black hole thermodynamics \cite{FGNP,BST,pranzetti} and on work on the large spin limit of spin foam models \cite{BN}. In \cite{FGNP}, it has been shown that in the context of the $\SU(2)$ Chern-Simons description of black holes in loop quantum gravity \cite{EPN,ENPP}, it is possible to recover the Bekenstein-Hawking formula for the entropy when $\gamma=\pm\mathrm{i}$. We find these relationships very interesting and encouraging, and suspect that they will clarify and indicate how to build a quantum theory based on the complex variables.

\section*{Aknowledgements}

\noindent We thank Sergei Alexandrov for very accurate comments and a careful reading of the manuscript, as well as Alejandro Perez and Simone Speziale. MG is supported by the NSF Grant PHY-1205388 and the Eberly research funds of The Pennsylvania State University.

\appendix

\section{The Lie algebras $\boldsymbol{\su(2)}$, $\boldsymbol{\sl(2,\mathbb{C})}$, and $\boldsymbol{\su(1,1)}$}
\label{appendix:algebra}

\noindent Let us first introduce the two-dimensional traceless Hermitian Pauli matrices
\be
\sigma_1=
\begin{pmatrix}
0	&	1	\\
1	&     0 
\end{pmatrix},\qquad
\sigma_2=
\begin{pmatrix}
0	&	-\mathrm{i}	\\
\mathrm{i}	&     0 
\end{pmatrix},\qquad
\sigma_3=
\begin{pmatrix}
1	&	0	\\
0	&     -1 
\end{pmatrix},
\ee
which form a basis of the Lie algebra $\su(2)$. One choice of basis for $\sl(2,\mathbb{C})$ is given by the rotation generators $J_i=-\mathrm{i}\sigma_i/2$ and the boost generators $P_i=-\sigma_i/2=-\mathrm{i}J_i$, with $i\in\{1,2,3\}$. They satisfy the following commutation relations:
\be\label{so(eta) basis}
[J_i,J_j]=\eps_{ij}^{~~k}J_k,\qquad[P_i,P_j]=-\eps_{ij}^{~~k}J_k,\qquad[P_i,J_j]=\eps_{ij}^{~~k}P_k.
\ee
One can see that the rotational algebra $\su(2)$ generated by the elements $J_i$ forms a subalgebra of the algebra $\sl(2,\mathbb{C})$. On the other hand, the subalgebra $\su(1,1)$ is generated by the elements $(J_3,P_1,P_2)$, and one can see from (\ref{so(eta) basis}) that their commutation relations are given by 
\be
[P_1,P_2]=-J_3,\qquad
[P_2,J_3]=P_1,\qquad
[J_3,P_1]=P_2.
\ee
In the literature, the Lie algebra $\su(1,1)$ is often defined as being generated by the three elements
\be\label{F generators}
F_0=
\f{1}{2}\begin{pmatrix}
-1	&	0	\\
0	&     1 
\end{pmatrix},\qquad
F_1=
\f{\mathrm{i}}{2}\begin{pmatrix}
0	&	1	\\
1	&     0 
\end{pmatrix},\qquad
F_2=
\f{1}{2}\begin{pmatrix}
0	&	1	\\
-1	&     0 
\end{pmatrix},
\ee
which satisfy the commutation relations
\be\label{su(1,1) brackets}
[F_1,F_2]=\mathrm{i}F_0,\qquad
[F_0,F_2]=\mathrm{i}F_1,\qquad
[F_0,F_1]=-\mathrm{i}F_2.
\ee
These generators are related to the previous ones through the map $(J_3,P_1,P_2)\longmapsto\mathrm{i}(F_0,F_1,F_2)$. Their commutation relations can be written in the more compact form
\be\label{su(1,1) bracket}
[F_i,F_j]=\mathrm{i}c_{ij}^{~~k}F_k,
\ee
where $i\in\{0,1,2\}$, and where the structure constants are given by
\be\label{su11 structure c}
c_{01}^{~~2}=-c_{10}^{~~2}=c_{20}^{~~1}=-c_{02}^{~~1}=c_{21}^{~~0}=-c_{12}^{~~0}=-1,
\ee
as can be easily verified by comparing (\ref{su(1,1) brackets}) with (\ref{su(1,1) bracket}).

Now, starting from the basis (\ref{so(eta) basis}), it is convenient to define a new basis $T_i^\pm$ as
\be
T_i^\pm\equiv\f{1}{2}(J_i\pm\mathrm{i}P_i),
\ee
whose generators realize two commuting copies of $\su(2)$, i.e. satisfy
\be
[T_i^\pm,T_j^\pm]=\eps_{ij}^{~~k}T_k^\pm,\qquad[T_i^+,T_j^-]=0.
\ee
Anti-symmetric bivectors $B^{IJ}$ form the adjoint representation of $\so(3,1)$. The Hodge duality operator acts on them as
\be
\star B^{IJ}=\f{1}{2}\eps^{IJ}_{~~KL}B^{KL},
\ee
which implies that $\star^2=-\text{id}$. We can therefore split the space of bivectors into the direct sum of two eigenspaces associated to the eigenvalues $\pm\mathrm{i}$, and write
\be
B^{IJ}=B^+_iT^{+IJ}_i+B^-_iT^{-IJ}_i.
\ee
The action of the Hodge dual on the (anti) self-dual components is given by
\be
\star B^\pm=\pm\mathrm{i}B^\pm,
\ee
and the vector representation of $\so(3,1)$ that we use is
\be
T^{\pm IJ}_i=\f{1}{2}\Big(\eps^{0iIJ}\pm\mathrm{i}\big(\eta^{0I}\eta^{iJ}-\eta^{iI}\eta^{0J}\big)\Big),
\ee
where $\eta^{IJ}=\text{diag}(-1,1,1,1)$.

\section{Commutator of two Ashtekar-Barbero connections}
\label{appendix:poisson}

\noindent The existence of the connection $\mathbf{A}$ (\ref{complex conn}) is a crucial point in the construction of section \ref{sec:quantum theory}. We have argued that this connection, shifted with a suitable term proportional to the Gauss constraint to give (\ref{asd}),  is canonically conjugated to the electric field $E$ (up to a global multiplicative factor). This argument relies on the fact that the shifted connection (\ref{asd}) corresponds to the self-dual or anti self-dual part of the initial $\sl(2,\mathbb{C})$ connection, which is itself conjugated to $E$. The non-trivial statement is that all the connections of the family (\ref{A and E}) (for any value of $\gamma$, either complex or real) are commutative. This can be proved by following the same reasoning as in the four-dimensional case and using properties of the components $\omega^i_a$ once the second class constraints are solved.

Due to the expression (\ref{asd}) where $\omega^i_a$ depends only on the variable $E^a_i$, showing that $\{A^i_a(x),A^j_b(y)\}=0$ reduces to the problem of showing that
\be
\{\omega_a^{(0)i}(x),\omega_b^j(y)\}+\{\omega_a^i(x),\omega_b^{(0)j}(y)\}=0,
\ee
for $a,b\in\{1,2\}$, $i,j\in\{1,2,3\}$ and $x,y\in\Sigma$, which in turn can be written as the condition
\be\label{crochetAG}
\{A_a^i(x),\omega_b^j(y)\}+\{\omega_a^i(x),A_b^j(y)\}=0,
\ee
since $\omega^i_a$ commutes with itself. To avoid a direct calculation of this relation, we proceed as in four dimensions, and look for a generating functional $W[E]$ depending on the variable $E$ only and such that $\omega_a^i(x)=\delta W[E]/\delta E^a_i(x)$. If this object exists, then the condition (\ref{crochetAG}) reduces to an integrability condition and follows immediately due to the fact that
\ba
\{A_a^i(x),\omega_b^j(y)\}+\{\omega_a^i(x),A_b^j(y)\}&=&\gamma\delta^2(x-y) 
\left(\frac{\delta\omega_a^i(x)}{\delta E^c_k(x)}\frac{\delta A_b^j(y)}{\delta A_c^k(y)}
-\frac{\delta A_a^i(x)}{\delta A_c^k(x)}\frac{\delta\omega_b^j(y)}{\delta E^c_k(y)}\right)\nonumber\\
&=&\gamma\delta^2(x-y) \left( \frac{\delta^2 W}{\delta E^a_i(x) \delta E^b_j(x)} - \frac{\delta^2 W}{\delta E^b_j(x) \delta E^a_i(x)} \right)\nonumber\\
&=&0.
\ea
For the generating functional we take
\be
W[E]=\int\de^2x\,E^a(x)\cdot\omega_a(x),
\ee
exactly as in four dimensions, with the difference that now $\omega^i_a$ is given by (\ref{sol omega}):
\be
\omega_a=u\time\partial_au+\frac{1}{|E|}\eps_{ab}(E^b\cdot\partial_c E^c)x.
\ee
For the functional $W$ to be such that $\omega_a^i(x)=\delta W[E]/\delta E^a_i(x)$, it should satisfy
\be
\int\de^2x\ E^a(x)\cdot\delta\omega_a(x)=0
\ee
for any variation $\delta\omega$ of $\omega$, which implies that
\be
\delta W=\int\de^2x\,(\delta E^a\cdot\omega_a+E^a\cdot\delta\omega_a)=
\int\de^2x\,\delta E^a\cdot\omega_a.
\ee
For this to be true, we assume that the spatial slice $\Sigma$ has no boundaries. The proof uses the fact that $\omega^i_a$ satisfies
\be\label{properties of omega}
\partial_aE^a-\omega_a\time E^a=0,\qquad\partial_au-\omega_a\time u=0.
\ee
Therefore, we have
\ba
\int\de^2x\,E^a\cdot\delta\omega_a&=&\int\de^2x\,E^a\cdot\left[\delta(u\time\partial_au)+
\delta\left(\frac{1}{|E|}\eps_{ab}E^b\cdot\partial_cE^cx\right)\right]\nonumber\\
&=&\int\de^2x\,E^a\cdot\left(\delta u\time\partial_au+u\time\partial_a\delta u+\frac{1}{|E|}\eps_{ab}E^b\cdot\partial_cE^c\delta x\right),\label{blabla}
\ea
where we used the fact that $x\cdot E^a=0$. The first two terms between parenthesis above can now be written as follows:
\ba
\int\de^2x\,\delta u\cdot\big(\partial_au\time E^a-\partial_aE^a\time u-E^a\time\partial_au\big)
&=&\int\de^2x\,\delta u\cdot\big(2(\omega_a\time E^a)u-(\omega_a\cdot u)E^a\big)\nonumber\\
&=&-\int\de^2x\,\delta u\cdot E^a(\omega_a\cdot u)\nonumber\\
&=&-\int\de^2x\,\frac{1}{|E|}E^a\cdot\delta x\eps_{ab}E^b\cdot\partial_cE^c,
\ea
from which (\ref{blabla}) vanishes as announced. Notice that from the first to the second line we used the properties (\ref{properties of omega}) to replace the derivatives $\partial_a E^a$ and $\partial_a u$ by expressions involving $\omega$.


\begin{thebibliography}{99}

\bibitem{barbero} J. F. Barbero,
Real Ashtekar variables for Lorentzian signature spacetimes,
Phys. Rev. \textbf{D 51} 5507 (1995).

\bibitem{immirzi} G. Immirzi,
Real and complex connections for canonical gravity,
Class. Quant. Grav. \textbf{14} L177 (1997), \texttt{arXiv:gr-qc/9612030}.

\bibitem{complexashtekar} A. Ashtekar,
New variables for classical and quantum gravity,
Phys. Rev. Lett. \textbf{57} 2244 (1986).

\bibitem{rovelli-thiemann} C. Rovelli and T. Thiemann,
The Immirzi parameter in quantum general relativity,
Phys. Rev. \textbf{D 57} 1009 (1998), \texttt{arXiv:gr-qc/9705059}.

\bibitem{menamarugan} G. A. Mena Marugan,
Extent of the Immirzi ambiguity in quantum general relativity,
Class. Quant. Grav. \textbf{19} L63 (2002), \texttt{arXiv:gr-qc/0203027}.

\bibitem{fermions1} L. Freidel, D. Minic and T. Takeuchi,
Quantum gravity, torsion, parity violation and all that,
Phys. Rev. \textbf{D 72} 104002 (2005), \texttt{arXiv:hep-th/0507253}.

\bibitem{fermions2} S. Mercuri,
Fermions in Ashtekar-Barbero connections formalism for arbitrary values of the Immirzi parameter,
Phys. Rev. \textbf{D 73} 084016 (2006), \texttt{arXiv:gr-qc/0601013}.

\bibitem{fermions3} A. Perez and C. Rovelli,
Physical effects of the Immirzi parameter,
Phys. Rev. \textbf{D 73} 044013 (2006), \texttt{arXiv:gr-qc/0505081}.

\bibitem{taveras-yunes} V. Taveras and N. Yunes,
The Barbero-Immirzi parameter as a scalar field: K-inflation from loop quantum gravity?,
Phys. Rev. \textbf{D 78} 064070 (2008), \texttt{arXiv:0807.2652 [gr-qc]}.

\bibitem{fermions4} M. Bojowald and R. Das,
Canonical gravity with fermions,
Phys. Rev. \textbf{D 78} 064009 (2008), \texttt{arXiv:0710.5722 [gr-qc]}.

\bibitem{fermions5} S. Alexandrov,
Immirzi parameter and fermions with non-minimal coupling,
Class. Quant. Grav. \textbf{25} 145012 (2008), \texttt{arXiv:0802.1221 [gr-qc]}.

\bibitem{mercuri-taveras} S. Mercuri and V. Taveras,
Interaction of the Barbero--Immirzi field with matter and pseudo-scalar perturbations,
Phys. Rev. \textbf{D 80} 104007 (2009), \texttt{arXiv:0903.4407 [gr-qc]}.

\bibitem{mercuri-gamma} S. Mercuri,
Peccei--Quinn mechanism in gravity and the nature of the Barbero--Immirzi parameter,
Phys. Rev. Lett. \textbf{103} 081302 (2009), \texttt{arXiv:0902.2764 [gr-qc]}.

\bibitem{mercuri-gamma2} S. Mercuri,
A possible topological interpretation of the Barbero-Immirzi parameter,
(2009), \texttt{arXiv:0903.2270 [gr-qc]}.

\bibitem{mercuri-randono} S. Mercuri and A. Randono,
The Immirzi parameter as an instanton angle,
Class. Quant. Grav. \textbf{28} 025001 (2011), \texttt{arXiv:1005.1291 [hep-th]}.

\bibitem{dittrich-ryan2} B. Dittrich and J. P. Ryan,
On the role of the Barbero-Immirzi parameter in discrete quantum gravity,
(2012), \texttt{arXiv:1209.4892 [gr-qc]}.

\bibitem{GN} M. Geiller and K. Noui,
A note on the Holst action, the time gauge, and the Barbero-Immirzi parameter,
Gen. Rel. Grav. \textbf{}  (2013), \texttt{arXiv:1212.5064 [gr-qc]}.

\bibitem{rovelli-smolin-area-volume} C. Rovelli and L. Smolin,
Discreteness of area and volume in quantum gravity,
Nucl. Phys. \textbf{B 442} 593 (1995), \texttt{arXiv:gr-qc/9411005}.

\bibitem{ashtekar-lewandowski-area} A. Ashtekar and J. Lewandowski,
Quantum theory of geometry I: Area operators,
Class. Quant. Grav. \textbf{14} A55 (1997), \texttt{arXiv:gr-qc/9602046}.

\bibitem{ashtekar-lewandowski-volume} A. Ashtekar and J. Lewandowski,
Quantum theory of geometry II: Volume operators,
Adv. Theor. Math. Phys. \textbf{1} 388 (1998), \texttt{arXiv:gr-qc/9711031}.

\bibitem{holst} S. Holst,
Barbero's Hamiltonian derived from a generalized Hilbert-Palatini action,
Phys. Rev. \textbf{D 53} 5966 (1996), \texttt{arXiv:gr-qc/9511026}.

\bibitem{aleSF} A. Perez,
The spin foam approach to quantum gravity,
Living Rev. Relativity \textbf{16} 3 (2013), \texttt{arXiv:1205.2019 [gr-qc]}.

\bibitem{FGNP} E. Frodden, M. Geiller, K. Noui and A. Perez,
Black hole entropy from complex Ashtekar variables,
(2012), \texttt{arXiv:1212.4060 [gr-qc]}.

\bibitem{BST} N. Bodendorfer, A. Stottmeister and A. Thurn,
Loop quantum gravity without the Hamiltonian constraint,
Class. Quant. Grav. \textbf{30} 082001 (2013), \texttt{arXiv:1203.6525 [gr-qc]}.

\bibitem{pranzetti} D. Pranzetti,
Black hole entropy from KMS-states of quantum isolated horizons,
(2013), \texttt{arXiv:1305.6714 [gr-qc]}.

\bibitem{BN} N. Bodendorfer and Y. Neiman,
Imaginary action, spinfoam asymptotics and the 'transplanckian' regime of loop quantum gravity,
(2013), \texttt{arXiv:1303.4752 [gr-qc]}.

\bibitem{Witten 3d} E. Witten,
$2+1$-dimensional gravity as an exactly soluble system,
Nucl. Phys. B \textbf{311} 46 (1988).

\bibitem{GN2} M. Geiller and K. Noui,
Testing the imposition of the spin foam simplicity constraints,
Class. Quant. Grav. \textbf{29} 135008 (2012), \texttt{arXiv:1112.1965 [gr-qc]}.

\bibitem{Alekseev:1994pa} A. Y. Alekseev, H. Grosse and V. Schomerus,
Combinatorial quantization of the Hamiltonian Chern-Simons theory,
Commun. Math. Phys. \textbf{172} 317 (1995), \texttt{arXiv:hep-th/9403066}.

\bibitem{Alekseev:1994au} A. Y. Alekseev, H. Grosse and V. Schomerus,
Combinatorial quantization of the Hamiltonian Chern-Simons theory. 2.,
Commun. Math. Phys. \textbf{174} 561 (1995), \texttt{arXiv:hep-th/9408097}.

\bibitem{BNR} E. Buffenoir, K. Noui and P. Roche,
Hamiltonian quantization of Chern-Simons theory with $\SL(2,\mathbb{C})$ group,
Class. Quant. Grav. \textbf{19} 4953 (2002), \texttt{arXiv:hep-th/0202121}.

\bibitem{Catherine} C. Meusburger and B. Schroers,
The quantisation of Poisson structures arising in Chern-Simons theory with gauge group $G\ltimes\mathfrak{g}^*$,
Adv. Theor. Math. Phys. \textbf{7} 1003 (2004), \texttt{arXiv:hep-th/0310218}.

\bibitem{wieland} W. Wieland,
Complex Ashtekar variables and reality conditions for Holst's action,
Annales H. Poincar\'e 1 (2011), \texttt{arXiv:1012.1738 [gr-qc]}.

\bibitem{alexandrov1} S. Alexandrov and D. V. Vassilevich,
Path integral for the Hilbert-Palatini and Ashtekar gravity,
Phys. Rev. \textbf{D 58} 124029 (1998), \texttt{arXiv:gr-qc/9806001}.

\bibitem{alexandrov2} S. Alexandrov,
SO(4,C)-covariant Ashtekar-Barbero gravity and the Immirzi parameter,
Class. Quant. Grav. \textbf{17} 4255 (2000), \texttt{arXiv:gr-qc/0005085}.

\bibitem{alexandrov4} S. Alexandrov and E. R. Livine,
SU(2) loop quantum gravity seen from covariant theory,
Phys. Rev. \textbf{D 67} 044009 (2003), \texttt{arXiv:gr-qc/0209105}.

\bibitem{GLNS} M. Geiller, M. Lachi\`eze-Rey, K. Noui and F. Sardelli,
A Lorentz-covariant connection for canonical gravity,
SIGMA \textbf{7} 083 (2011), \texttt{arXiv:1103.4057 [gr-qc]}.

\bibitem{GLN} M. Geiller, M. Lachi\`eze-Rey and K. Noui,
A new look at Lorentz-covariant loop quantum gravity,
Phys. Rev. \textbf{D 84} 044002 (2011), \texttt{arXiv:1105.4194 [gr-qc]}.

\bibitem{alexandrov5} S. Alexandrov,
On choice of connection in loop quantum gravity,
Phys. Rev. \textbf{D 65} 024011 (2001), \texttt{arXiv:gr-qc/0107071}.

\bibitem{alexandrov6} S. Alexandrov,
Hilbert space structure of covariant loop quantum gravity,
Phys. Rev. \textbf{D 66} 024028 (2002), \texttt{arXiv:gr-qc/0201087}.

\bibitem{dittrich-thiemann} B. Dittrich and T. Thiemann,
Are the spectra of geometrical operators in loop quantum gravity really discrete?,
J. Math. Phys. \textbf{50} 012503 (2009), \texttt{arXiv:0708.1721 [gr-qc]}.

\bibitem{giesel-thiemann} K. Giesel and T. Thiemann,
Algebraic quantum gravity (AQG) IV. Reduced phase space quantisation of loop quantum gravity,
(2007), \texttt{arXiv:0711.0119 [gr-qc]}.

\bibitem{DGKL} M. Domagala, K. Giesel, W. Kaminski and J. Lewandowski,
Gravity quantized,
Phys. Rev. \textbf{D 82} 104038 (2010), \texttt{arXiv:1009.2445 [gr-qc]}.

\bibitem{BST2} N. Bodendorfer, A. Stottmeister and A. Thurn,
On a partially reduced phase space quantisation of general relativity conformally coupled to a scalar field,
Class. Quant. Grav. \textbf{30} 115017 (2013), \texttt{arXiv:1203.6526 [gr-qc]}.

\bibitem{Bonzon Livine} V. Bonzom and E. Livine,
A Immirzi-like parameter for 3d quantum gravity,
Class. Quant. Grav. \textbf{25} 195024 (2008), \texttt{arXiv:0801.4241 [gr-qc]}.

\bibitem{urbantke} H. Urbantke,
On integrability properties of SU(2) Yang-Mills fields. I. Infinitesimal part,
J. Math. Phys. \textbf{25} 2321 (1984).

\bibitem{Sergey unpublished} S. Alexandrov,
(2012), unpublished.

\bibitem{mitra-rajaraman} P. Mitra and R. Rajaraman,
Gauge-invariant reformulation of an anomalous gauge theory,
Phys. Lett. B \textbf{225} 267 (1989).

\bibitem{samuel} J. Samuel,
Is Barbero's Hamiltonian formulation a gauge theory of Lorentzian gravity?,
Class. Quant. Grav. \textbf{17} L141 (2000), \texttt{arXiv:gr-qc/0005095}.

\bibitem{Achucarro-Townsend} A. Ach\'ucarro and P. K. Townsend,
A Chern-Simons action for three-dimensional anti-de Sitter supergravity theories,
Phys. Lett. B \textbf{180} 89 (1986).

\bibitem{Witten Jones} E. Witten,
Quantum field theory and the Jones polynomial,
Commun. Math. Phys. \textbf{121} 351 (1989).

\bibitem{ponzano-regge} G. Ponzano and T. Regge,
in \textit{Spectroscopy and group theoretical methods in physics},
ed. by F. Block (North Holland, 1968).

\bibitem{turaev-viro} V. G. Turaev and O. Y. Viro,
State sum invariants of 3-manifolds and quantum 6j-symbols,
Topology \textbf{31} 865 (1992), \texttt{arXiv:gr-qc/0402110}.

\bibitem{BC} J. W. Barrett and L. Crane,
Relativistic spin networks and quantum gravity,
J. Math. Phys. \textbf{39} 3296 (1998), \texttt{arXiv:gr-qc/9709028}.

\bibitem{EPR} J. Engle, R. Pereira and C. Rovelli,
Flipped spinfoam vertex and loop gravity,
Nucl. Phys. \textbf{B 798} 251 (2008), \texttt{arXiv:0708.1236 [gr-qc]}.

\bibitem{EPRL} J. Engle, E. R. Livine, R. Pereira and C. Rovelli,
LQG vertex with finite Immirzi parameter,
Nucl. Phys. \textbf{B 799} 136 (2008), \texttt{arXiv:0711.0146 [gr-qc]}.

\bibitem{livine-speziale1} E. R. Livine and S. Speziale,
Consistently solving the simplicity constraints for spinfoam quantum gravity,
Europhys. Lett. \textbf{81} 50004 (2008), \texttt{arXiv:0708.1915 [gr-qc]}.

\bibitem{livine-speziale2} E. R. Livine and S. Speziale,
A new spinfoam vertex for quantum gravity,
Phys. Rev. \textbf{D 76} 084028 (2007), \texttt{arXiv:gr-qc/0705.0674}.

\bibitem{FK} L. Freidel and K. Krasnov,
A new spin foam model for 4d gravity,
Class. Quant. Grav. \textbf{25} 125018 (2008), \texttt{arXiv:gr-qc/0708.1595 [gr-qc]}.

\bibitem{Noui:2004iy} K. Noui and A. Perez,
Three dimensional loop quantum gravity: Physical scalar product and spin foam models,
Class. Quant. Grav. \textbf{22} 1739 (2005), \texttt{arXiv:gr-qc/0402110}.

\bibitem{Noui:2004iz} K. Noui and A. Perez,
Three dimensional loop quantum gravity: Coupling to point particles,
Class. Quant. Grav. \textbf{22} 4489 (2005), \texttt{arXiv:gr-qc/0402111}.

\bibitem{AGN} S. Alexandrov, M. Geiller and K. Noui,
Spin foams and canonical quantization,
Sigma \textbf{8} 055 (2012), \texttt{arXiv:1112.1961 [gr-qc]}.

\bibitem{Witten analytic} E. Witten,
Analytic continuation of Chern-Simons theory,
(2010), \texttt{arXiv:1001.2933 [hep-th]}.

\bibitem{Cath and I} C. Meusburger and K. Noui,
The Hilbert space of 3d gravity: quantum group symmetries and observables,
Adv. Theor. Math. Phys. \textbf{14} 1651 (2010), \texttt{arXiv:0809.2875 [gr-qc]}.

\bibitem{Cath and I2} C. Meusburger and K. Noui,
Combinatorial quantization of the Euclidean torus universe,
Nuclear Physics B \textbf{841} 463 (2010), \texttt{arXiv:1007.4615 [gr-qc]}.

\bibitem{Freidel Livine} L. Freidel and E. Livine,
Spin networks for non-compact groups,
Math. Phys. \textbf{44} 1322 (2003), \texttt{arXiv:hep-th/0205268}.

\bibitem{BTZ-SF} J. M. Garcia-Islas,
BTZ black hole entropy: A spin foam model description,
Class. Quant. Grav. \textbf{25} 245001 (2008), \texttt{arXiv:0804.2082 [gr-qc]}.

\bibitem{ale-amb} A. Perez,
On the regularization ambiguities in loop quantum gravity,
Phys. Rev. \textbf{D 73} 044007 (2006), \texttt{arXiv:gr-qc/0509118}.

\bibitem{Bonzom:2011jv} V. Bonzom and A. Laddha,
Lessons from toy-models for the dynamics of loop quantum gravity,
Sigma \textbf{8} 055 (2012), \texttt{arXiv:1110.2157 [gr-qc]}.

\bibitem{conrady} F. Conrady and J. Hnybida,
A spin foam model for general Lorentzian 4-geometries,
Class. Quant. Grav. \textbf{27} 185011 (2010), \texttt{arXiv:1002.1959 [gr-qc]}.

\bibitem{BGNY} J. Ben Achour, M. Geiller, K. Noui and C. Yu,
Spectra of geometric operators in three-dimensional LQG: From discrete to continuous
(2013), \texttt{arXiv: [gr-qc]}.

\bibitem{freidel-livine-rovelli} L. Freidel, E. R. Livine and C. Rovelli,
Spectra of length and area in 2+1 Lorentzian loop quantum gravity,
Class. Quant. Grav. \textbf{20} 1463 (2003), \texttt{arXiv:gr-qc/0212077}.

\bibitem{Alexandrov:2010pg} S. Alexandrov,
The new vertices and canonical quantization,
Phys. Rev. \textbf{D 82} 024024 (2010), \texttt{arXiv:1004.2260 [gr-qc]}.

\bibitem{EPN} J. Engle, A. Perez and K. Noui,
Black hole entropy and SU(2) Chern-Simons theory,
Phys. Rev. Lett. \textbf{105} 031302 (2010), \texttt{arXiv:0905.3168 [gr-qc]}.

\bibitem{ENPP} J. Engle, K. Noui, A. Perez and D. Pranzetti,
Black hole entropy from an SU(2)-invariant formulation of type I isolated horizons,
Phys. Rev. \textbf{D 82} 044050 (2010), \texttt{arXiv:1006.0634 [gr-qc]}.

\end{thebibliography}
\end{document}